\pdfoutput=1
\documentclass[article,aps,nofootinbib,twocolumn,superscriptaddress]{revtex4-1}
\input epsf
\usepackage{graphics}
\usepackage{amsmath}
\usepackage{amssymb}
\usepackage{bm}
\usepackage{braket}
\usepackage{booktabs}
\usepackage{subfigure}
\usepackage{subfloat}
\usepackage{float}
\usepackage[usenames,svgnames]{xcolor}
\usepackage{hyperref}

\hypersetup{
    colorlinks=true,
    urlcolor=SteelBlue,
    linkcolor=red,
    citecolor=blue,
}

\usepackage{color}
\usepackage{dcolumn}
\usepackage{hyphenat}

\def\be{\begin{equation}}
\def\ee{\end{equation}}
\def\ba{\begin{eqnarray}}
\def\ea{\end{eqnarray}}

\usepackage{graphicx}

\begin{document}

\title{Echoes from the scattering of wavepackets on wormholes}

\author{Jos\'e T. G\'alvez Ghersi}
\email{joseg@sfu.ca}
\author{Andrei V. Frolov}
\email{frolov@sfu.ca}
\author{David A. Dobre}
\email{ddobre@sfu.ca}
\affiliation{Department of Physics, Simon Fraser University,\\
 8888 University Drive, Burnaby, British Columbia V5A 1S6, Canada}
\date{\today}

\begin{abstract}
It has been recently shown that observing pulses isolated from the gravitational radiation transient (also known as echoes) would prove the existence of exotic compact objects (ECOs). Many features of the ringdown signal can be reproduced by simulating a scattering problem instead of the full coalescence of ECOs. In this paper, we study the dynamics of scalar and tensor wavepackets colliding against a spherically symmetric Morris-Thorne wormhole. Our aim is to extract the features of the time-dependent scattering solutions inside and outside the effective potential cavity in addition to their asymptotic behavior. Using the geometrical optics approximation, we show that the amplitude of the echoes is only large enough in a narrow bandwidth of frequency space. Additionally, we show that the cavity modifies the polarization of the asymptotic gravitational wave solutions. The computer code used to produce these results is publicly available for further applications, including scattering and accretion processes.      
\end{abstract}

\maketitle

\section{Introduction}
The era of gravitational wave (GW) astronomy \citep{Abbott:2016blz, Abbott:2016nmj} has begun. GW spectroscopy, in contrast to its atomic counterpart, allows us to characterize strong gravitational interactions in their radiative regime. In this new range of frequencies, it is now possible to explore the role of dynamical gravitational degrees of freedom in a wide range of astrophysical \citep{Frolov:2017asg, Cardoso:2016rao} and cosmological \citep{Krauss989, Ade:2018gkx} phenomena. 

The prolonged absence of observational evidence confirming the dynamical properties of spacetime has motivated a plethora of conjectures about the behaviour of gravity within and beyond \citep{Clifton:2011jh, Taliotis:2012sx, Joyce:2014kja} classical General Relativity (GR). The potential existence of exotic compact objects (ECOs) sourced by quantum effects on gravity \citep{PhysRevLett.61.1446, Almheiri:2012rt, Mazur:2001fv} (such as wormholes, firewalls and gravastars) has captured the attention of many recent efforts \citep{Kokkotas:1995av, Tominaga:1999iy, Ferrari:2000sr, Cardoso:2016oxy, Abedi:2016hgu, Price:2017cjr, Maselli:2017tfq, Volkel:2017kfj, Conklin:2017lwb, Mark:2017dnq, Bueno:2017hyj, Volkel:2018hwb, Abedi:2018npz, Wang:2018mlp, Correia:2018apm, Testa:2018bzd, Urbano:2018nrs, Chen:2019hfg}. The primary claim is that the detection of a train of ``echoes'' isolated from the main transient of GW and with generically large amplitudes would be clear evidence of ECOs. It is, therefore, necessary to understand (i) the mechanisms behind the production of echoes and (ii) the intensity and spectrum of the outgoing wavelets compared to the GW transient in the most straightforward possible setup. In this paper, we explore the generation of echoes by colliding wavepackets of scalar and tensor radiation against a traversable spherically symmetric wormhole \citep{Visser:1989kh}. We find that such a wormhole behaves just like a Fabry-Perot cavity and shares common properties with the effective potential cavities made by other ECOs, like gravastars and firewalls. Additionally, the main features of the outgoing pulses are similar to the ringdown signals expected from the coalescence of ECOs.

Here we consider a simplified wormhole configuration made by the junction of two Schwarzschild geometries of equal masses at $r_0>2M$, widely known as the Morris-Thorne wormhole \citep{Cardoso:2016oxy, doi:10.1119/1.15620}. In this case, the symmetry of the centrifugal barriers at $r=3M$ on each side of the throat allows us to find the reflection and transmission coefficients of the cavity. Hence, it is possible to reconstruct the spectral shape of the outgoing pulse using the geometrical optics approximation. Nevertheless, this approximation predicts an exponential decay of the subsequent higher order reflections, which appears instead as a power law in the full solution of the scattering problem. Thus, the excitation of quasinormal modes (QNMs) is the only cause for the presence of echoes in the time evolving profile. These modes are sourced by a sequence of internal reflections inside the potential cavity and then propagate throughout the surface of the maximal potential energy spheres (i.e., the ``edges'' of the potential barriers), while radiating energy to the exterior. QNMs of the Schwarzschild solution have been extensively studied and reproduced in various analytic and numerical simulations \citep{Chandrasekhar:1975zza, PhysRevD.46.4179}; thus it is easy to identify their characteristic frequencies in the spectrum of outgoing pulses. We also present the full scattering solution both inside and outside the wormhole cavity in detail, along with the energy fluxes and the asymptotic solutions for the principal spherical modes of a scalar (and tensor) wavepacket. In addition to this, we find the width and frequency intervals contained in the incident wavepackets for which the outgoing wavelets have maximal amplitudes.

In the scattering of gravitational radiation, we observe that the shape of the potential walls is different for the even and odd polarization modes. Due to this difference, we find that the echoes from an unpolarized ingoing Gaussian wavepacket obtain a small net polarization, even when there is no distinction in their spectral content. Our computer code is optimized to solve both scalar accretion and scattering problems, and is publicly available at \url{https://github.com/andrei-v-frolov/accretion/tree/wormhole}.

The layout for this paper is as follows: in section \ref{sec:scalar}, we review the scattering problem of scalar waves starting by a quick overview of the dispersion of a Gaussian pulse by a Schwarzschild black hole. The main point of this section is to calculate the transmission and reflection coefficients of each of the centrifugal barriers constituting the resonant cavity, formed in the case of a wormhole. Our results show a frequency ``sweet spot'' such that the incident pulse is not fully reflected nor fully transmitted by the cavity, favouring multiple internal reflections that source the QNMs.  Furthermore, we solve the problem of scattering by a wormhole directly using the same ingoing Gaussian wavepacket, and then we compare the Fourier transform of this solution with the pulse reconstructed following the geometrical optics approximation. We find that the approximate reconstruction matches the full solution, up to the peaks due to the QNM frequencies. Likewise, we evaluate the amplitude of each of the echoes as a function of the width of the initial gaussian waveform, finding that a single width of the incident pulse maximizes the amplitude of all the echoes. In Section \ref{sec:tensor}, we extend the results in the previous section to a Gaussian pulse of tensor fluctuations of the metric by following the even and odd decomposition of the tensor modes introduced by Regge, Wheeler and Zerilli in \citep{Regge:1957td, PhysRevD.2.2141, PhysRevD.5.2419, PhysRevD.5.2439}. Our results can be recasted in terms of the usual asymptotic polarization modes $h_+$ and $h_{\times}$, known as the perturbations of a flat metric. In this section, we also observe that the echoes from an unpolarized incident mode obtain a small piecewise net polarization in spite of the spherical symmetry of the scattering target. We conclude with a discussion in Section \ref{sec:conclusions}.

\section{Scattering of scalar wavepackets}\label{sec:scalar}
In this section, we solve the scattering of a Gaussian wavepacket by a spherically symmetric wormhole. To do so, we will first review the dispersion by the centrifugal barrier of a spherically symmetric black hole in order to find the properties of the potential cavity.
\subsection{Scattering by a Schwarzschild black hole}\label{subsec:bh} 
Our primary objective is to study the dynamics of scalar and tensor wavepackets scattering a Misner-Thorne wormhole at all points. Thus, we first review the dispersion of scalar waves by a Schwarzschild black hole, thoroughly studied in \citep{doi:10.1063/1.522949, Sanchez:1976xm, Sanchez:1977si, Sanchez:1977vz}, wherein the collision against each of the two potential walls (constituting the effective potential cavity formed by a wormhole) is studied in full detail. The dynamics of the scattering problem is found by solving the equation of motion for a test scalar field
\begin{equation}
\Box\Phi=0,\label{eq:scalar_eq_mov} 
\end{equation}
where $\Box\equiv g^{\alpha\beta}\nabla_{\alpha}\nabla_{\beta}$ is the standard d'Alembertian in a curved background. Here $g_{\alpha\beta}$ is the metric tensor in a spherically symmetric Schwarzschild-like static spacetime
\begin{equation}
g_{\alpha\beta}=-f(r)\delta^t_{\alpha}\delta^t_{\beta}+\frac{1}{f(r)}\delta^r_{\alpha}\delta^r_{\beta}+r^2\left(\delta^{\theta}_{\alpha}\delta^{\theta}_{\beta}+\sin^2\theta\delta^{\phi}_{\alpha}\delta^{\phi}_{\beta}\right).
\label{eq:Schwarzschild}
\end{equation}
\begin{figure}[t]
\centering
\includegraphics[width=.45\textwidth]{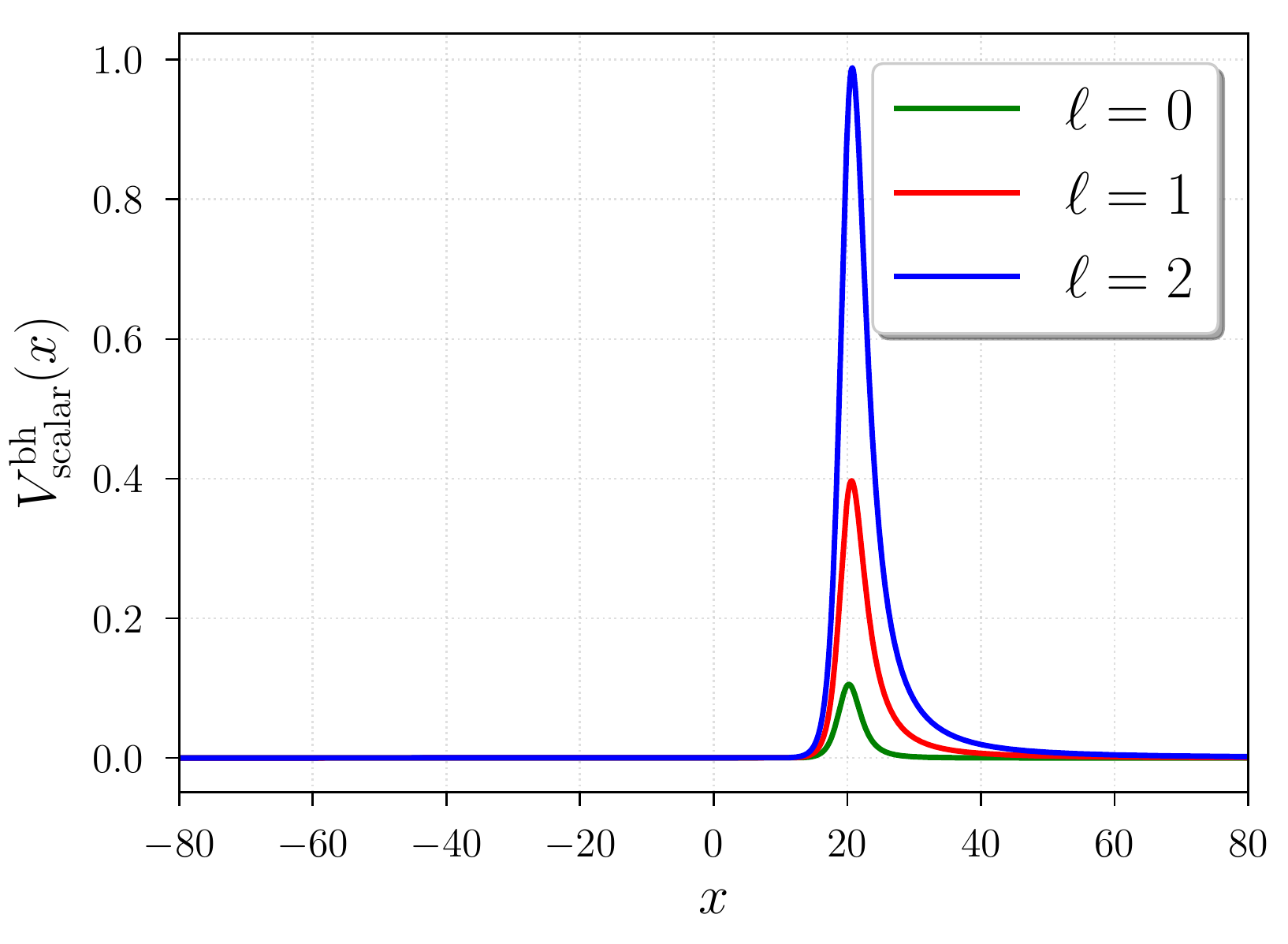}
\caption{\label{fig:Potential_BH} Effective potential for the spherical modes $\mathcal{U}^{\mathrm{bh}}_{20}(x,t)$ scattered by a Schwarzschild black hole, growing with $\ell$. The wall acts as a barrier transparent to certain frequencies above a transmissivity threshold and reflective for lower frequencies.}
\end{figure}
It is convenient to introduce the tortoise coordinate $x$:
\begin{equation}
x\equiv\displaystyle{\int_{r_0}^r\frac{dr}{f(r)}},
\label{eq:tortoise}
\end{equation} 
In the case of the Schwarzschild metric $f(r)=1-r_g/r$ the last expression yields
\begin{equation}
x=r-r_0+r_g\ln\left(\frac{r-r_g}{r_0-r_g}\right),
\label{eq:tortoise2}
\end{equation} 
for $r_g<r<+\infty$ and $r_0>r_g$, where $r_g=2M$ is the usual Schwarzschild radius. By direct evaluation, we see that $r=r_0$ corresponds to $x=0$, the horizon $r=2M$ maps into $x\rightarrow-\infty$ and $r\rightarrow+\infty$ is $x\rightarrow+\infty$. In our numerical routine, we invert \eqref{eq:tortoise2} to get $r\equiv r(x)$ (see the appendix A, subsection 6 in \citep{Frolov:2017asg} for more details). In tortoise coordinates, we can decompose the scalar field in spherical harmonics 
\begin{equation}
\Phi(x,t)=\frac{1}{r(x)}\sum_{\ell,m=0}\mathcal{U}^{\mathrm{bh}}_{\ell m}(x,t)Y_{\ell m}(\theta,\phi),
\label{eq:ylm_decomp}
\end{equation}
in that way we can rewrite \eqref{eq:scalar_eq_mov} as
\begin{equation}
\left[-\partial_t^2+\partial_x^2-V^{\mathrm{bh}}_{\mathrm{scalar}}(x)\right]\mathcal{U}^{\mathrm{bh}}_{\ell m}(x,t) = 0,
\label{eq:wave_scalar}
\end{equation}
and the effective potential $V^{\mathrm{bh}}_{\mathrm{scalar}}(x)$ is given by
\begin{equation}
V^{\mathrm{bh}}_{\mathrm{scalar}}(x) = \left(1-\frac{r_g}{r(x)}\right)\left[\frac{\ell(\ell+1)}{r(x)^2}+\frac{r_g}{r(x)^3}\right].
\end{equation}
After rearranging the variables, the equation of motion of the spherical modes is now written in its traditional linear waveform. In Fig.~\ref{fig:Potential_BH}, we observe the growth of the potential barrier with the angular momentum number $\ell$. The potential wall does not vanish for the monopole ($\ell=0$) due to the extra term proportional to $r^{-3}$ appearing after the coordinate change, which replaces the radial damping in the original Schwarzschild coordinates $(t,r)$. Such a term becomes subdominant for all $\ell\geq 1$. Intuitively, it is reasonable to expect that the modes with frequency above a given threshold (related with the thickness of the wall) can cross the barrier, while reflecting the lower frequency modes. 

Now we setup the scattering problem for one of the spherical modes ($\mathcal{U}^{\mathrm{bh}}_{20}$, the quadrupole) with the following initial conditions corresponding to an ingoing Gaussian wavepacket 
\begin{equation}
\mathcal{U}^{\mathrm{bh}}_{20}(x,0)=\exp\left(\frac{(x-x_0)^2}{2\sigma^2}\right)~,~ \partial_t\mathcal{U}^{\mathrm{bh}}_{20}\bigg{|}_{t=0}=\partial_x \mathcal{U}^{\mathrm{bh}}_{20}(x,0),
\label{eq:scalar_init_cond}
\end{equation}

\begin{figure}[t!]
\centering
\includegraphics[width=.45\textwidth]{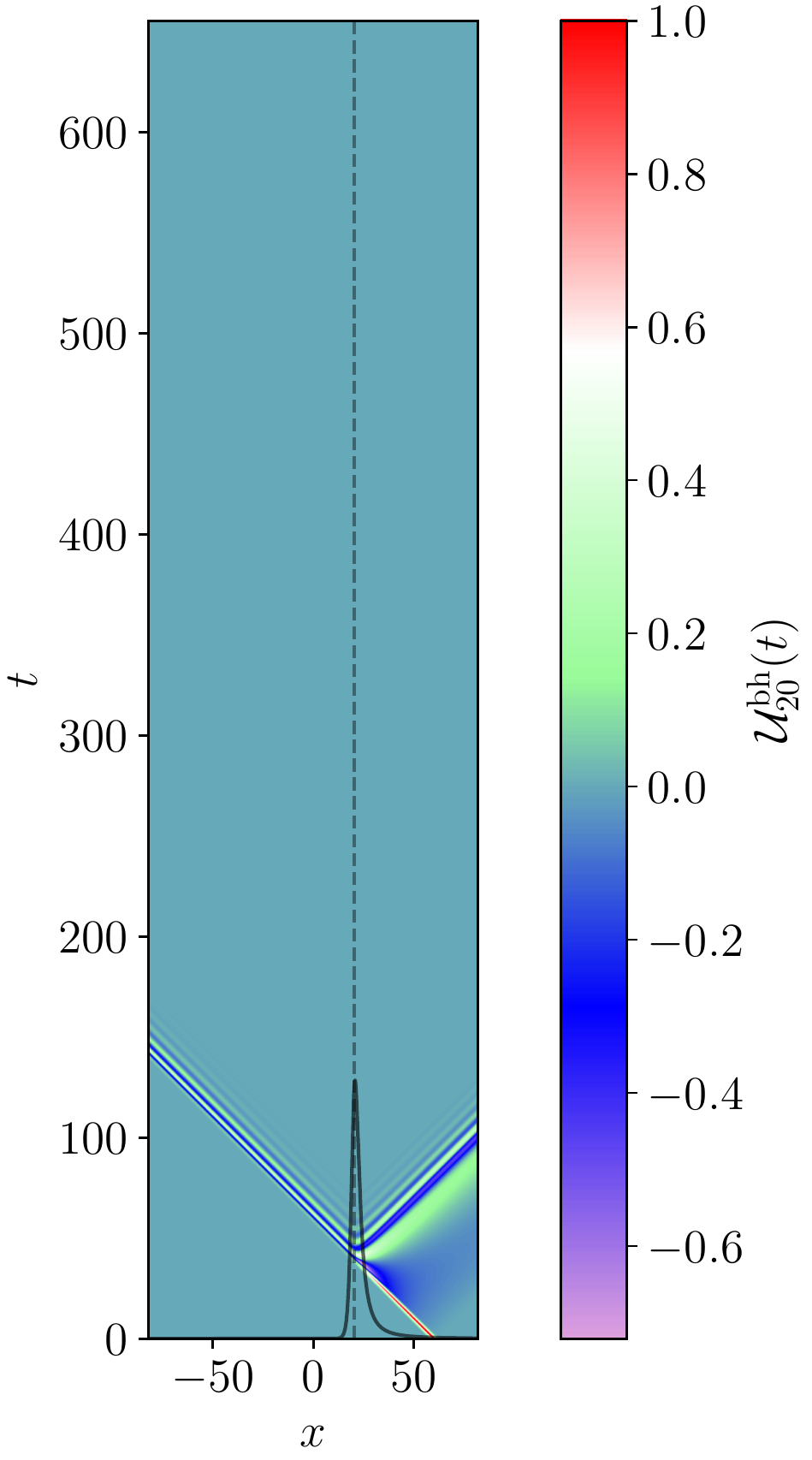}
\caption{\label{fig:bh_sol} Dispersion of the ingoing Gaussian wavepacket $\mathcal{U}^{\mathrm{bh}}_{20}$ by the potential barrier (plotted in black) showing the incident, reflected and transmitted pulses, it is possible notice the ringing of the reflected solution due to the quasinormal modes.}
\end{figure}

\begin{figure}[t]
\centering
\includegraphics[width=.45\textwidth]{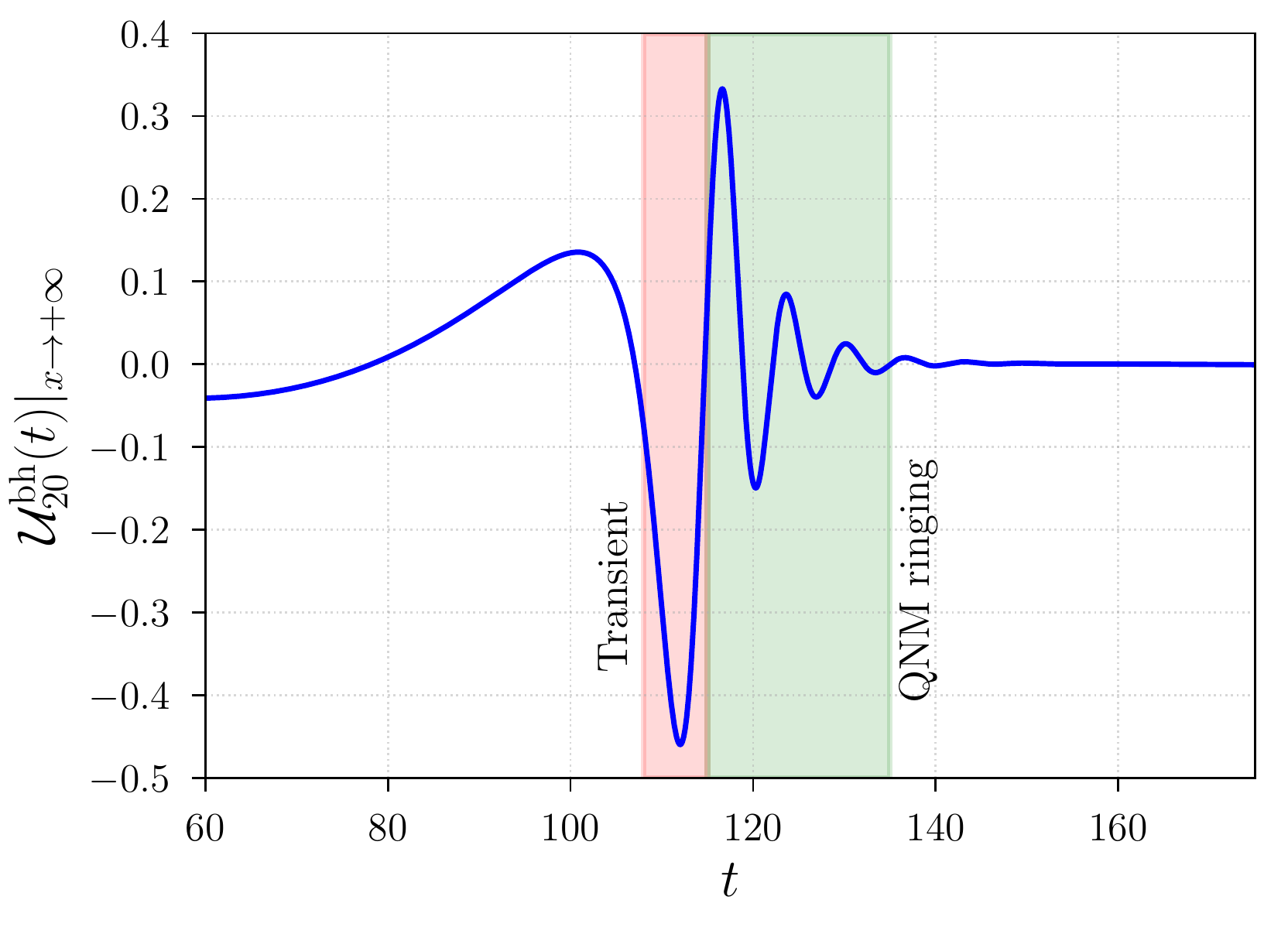}
\caption{\label{fig:asympt_sol} Asymptotic solution for the quadrupole mode $\mathcal{U}^{\mathrm{bh}}_{20}(x,t)$ by direct evaluation of the results in Fig.~\ref{fig:bh_sol}. The reflected signal shows its maximum peak and the posterior ringing due to QNMs.}
\end{figure}

\begin{figure}[t]
\centering
\includegraphics[width=.45\textwidth]{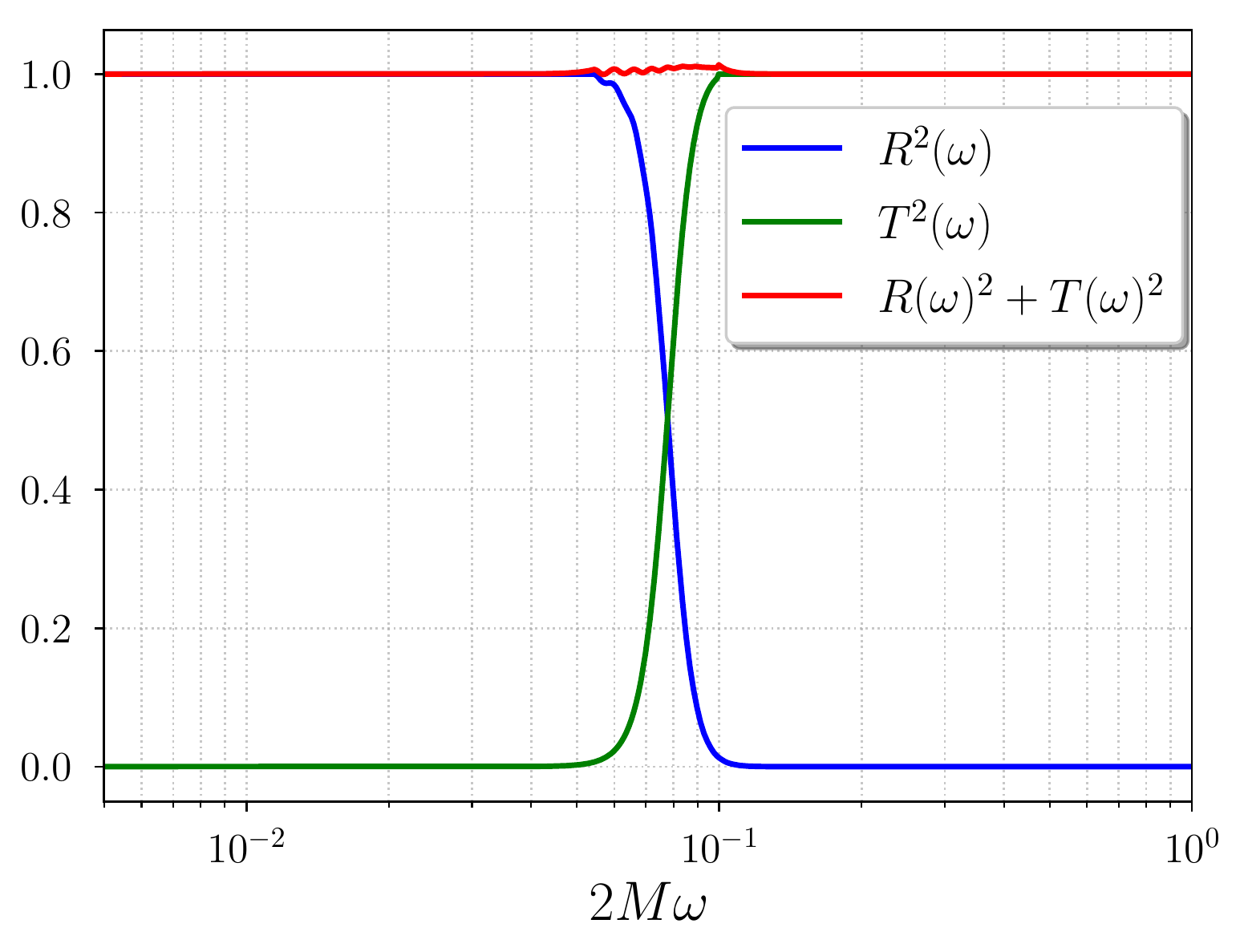}
\caption{\label{fig:RT_scalar} Reflection and transmission coefficients as a function of frequency $(\omega)$, the identity $R^2+T^2=1$ is satisfied with an error smaller than 1\%.}
\end{figure}

After fixing the values of the width to be $\sigma=0.9185r_g$, the initial position of the Gaussian at $x_0=60.0r_g$, $r_0=20.0r_g$ and the initial conditions in \eqref{eq:scalar_init_cond}, we show the time-dependent solution of \eqref{eq:wave_scalar} in Fig.~\ref{fig:bh_sol}, where we distinguish the incident, transmitted and reflected parts of the solution. It is important to observe the absence of spurious late time reflections and interferences due to the implementation of perfectly matching layers (PMLs) in the outermost regions of our simulation box (see the details of our setup for PMLs in \citep{Frolov:2017asg}). We observe the main features of the reflected signal in Fig.~\ref{fig:asympt_sol}, where the asymptotic behavior of the signal shows a sharp transient as a consequence of the collision against the potential wall, and the ringing of quasinormal modes occurring right after the reflection in agreement with \citep{Petrich:1985csm}. 

It is now possible to evaluate the reflection and transmission coefficients of each of the potential walls depicted in Fig.~\ref{fig:Potential_BH}. To do so, we compute the one dimensional Fourier transform of the incident $\tilde{\mathcal{U}}_{20}^{\mathrm{inc}}(\omega)=\mathcal{F}[\mathcal{U}^{\mathrm{bh}}_{20}(0,x)]$, reflected $\tilde{\mathcal{U}}_{20}^{\mathrm{ref}}(\omega)=\mathcal{F}[\mathcal{U}^{\mathrm{bh}}_{20}(t,+\infty)]$ and transmitted $\tilde{\mathcal{U}}_{20}^{\mathrm{trans}}(\omega)=\mathcal{F}[\mathcal{U}^{\mathrm{bh}}_{20}(t,-\infty)]$ from the solved scattering modes in order to define
\begin{equation}
R(\omega)\equiv \frac{||\tilde{\mathcal{U}}_{20}^{\mathrm{ref}}(\omega)||}{||\tilde{\mathcal{U}}_{20}^{\mathrm{inc}}(\omega)||}~,~T(\omega)\equiv \frac{||\tilde{\mathcal{U}}_{20}^{\mathrm{trans}}(\omega)||}{||\tilde{\mathcal{U}}_{20}^{\mathrm{inc}}(\omega)||}
\label{eq:ref_and_trans}
\end{equation}
\begin{figure}[t]
\centering
\includegraphics[width=.4\textwidth]{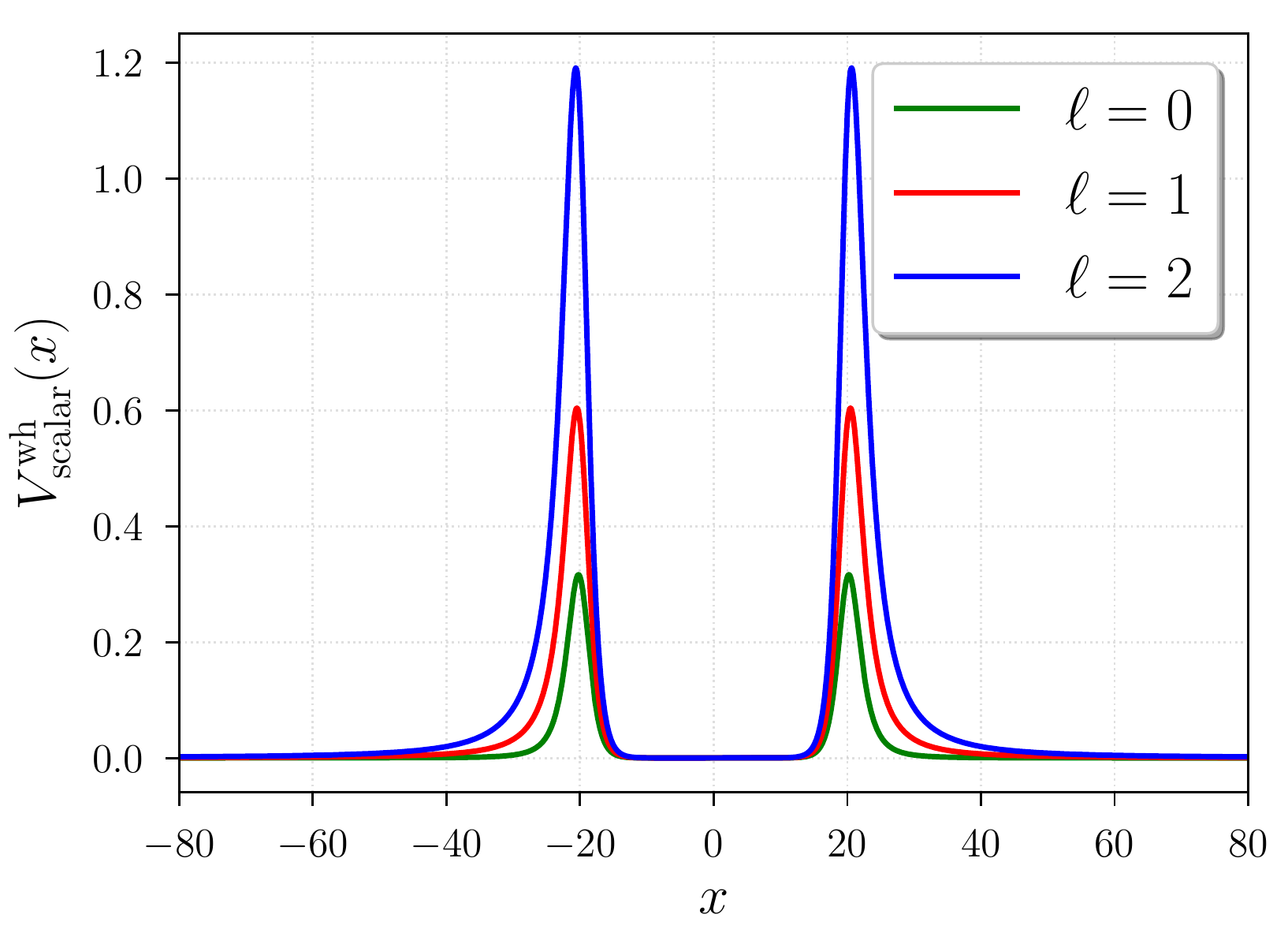}
\caption{\label{fig:potential_wh} Effective potential cavity for the wormhole, we observe the growth of the barriers with the angular momentum number $\ell$.}
\end{figure}

as the transmission and reflection coefficients of an arbitrary potential barrier, respectively. In Fig.~\ref{fig:RT_scalar}, we plot the squares of these coefficients as functions of frequency observing that the identity $R^2+T^2=1$ is only approximately met because of the small contributions coming from the QNMs frequency peaks in both the transmitted and reflected solutions. The shape of both the transmissivity and reflectivity curves is very similar to an hyperbolic tangent step function\footnote{This is not surprising after we consider the DeWitt approximation for the transmissivity \citep{DeWitt:1975ys,Frolov:1998wf}, which is precisely given by a step function.}, intersecting at $R^2=T^2=0.5$, as expected. Furthermore, it is crucial to notice from the last figure that it is only in a narrow band of frequencies where the amplitudes transmitted and reflected by the potential barrier are comparable. in the case of a wormhole, such a fact will be important in our analysis.  

\subsection{Scattering by a traversable wormhole}\label{subsec:wh} 
In this section, we study the dispersion of Gaussian wavepackets by a traversable wormhole, formed by the junction at $r_0>r_g$ of two Schwarzschild black hole solutions with equal mass. There is a discontinuity in $G_{\alpha\beta}$ such that at $r=r_0=20.0r_g$, any contracting congruence of geodesics in one side of the throat starts to expand in order to reach the other side, violating the weak energy condition. Such a configuration is known in the literature as the Morris-Thorne traversable wormhole \citep{doi:10.1119/1.15620}. It is important to mention that the Morris-Thorne solution slowly evolves into a black hole configuration, as shown in \citep{Wang:2018mlp}, and therefore, we should consider these results are only viable in intervals much shorter than the corresponding transition timescales. The dynamics of the scattering problem is still given by the solutions of \eqref{eq:scalar_eq_mov} following the same decomposition in spherical modes as in \eqref{eq:ylm_decomp}. Hence, the waveform of the equation of motion for the spherical modes is given by
\begin{equation}
\left[-\partial_t^2+\partial_x^2-V^{\mathrm{wh}}_{\mathrm{scalar}}(x)\right]\mathcal{U}_{\ell m}(x,t) = 0,
\label{eq:wave_scalar_wh}
\end{equation}
and the effective potential $V^{\mathrm{wh}}_{\mathrm{scalar}}(x)$ yields
\begin{equation}
V^{\mathrm{wh}}_{\mathrm{scalar}}\left(x\right) = \left(1-\frac{r_g}{r\left(|x|\right)}\right)\left[\frac{\ell(\ell+1)}{r\left(|x|\right)^2}+\frac{r_g}{r\left(|x|\right)^3}\right].
\end{equation}   
\begin{figure}[t!]
\centering
\includegraphics[width=.3\textwidth]{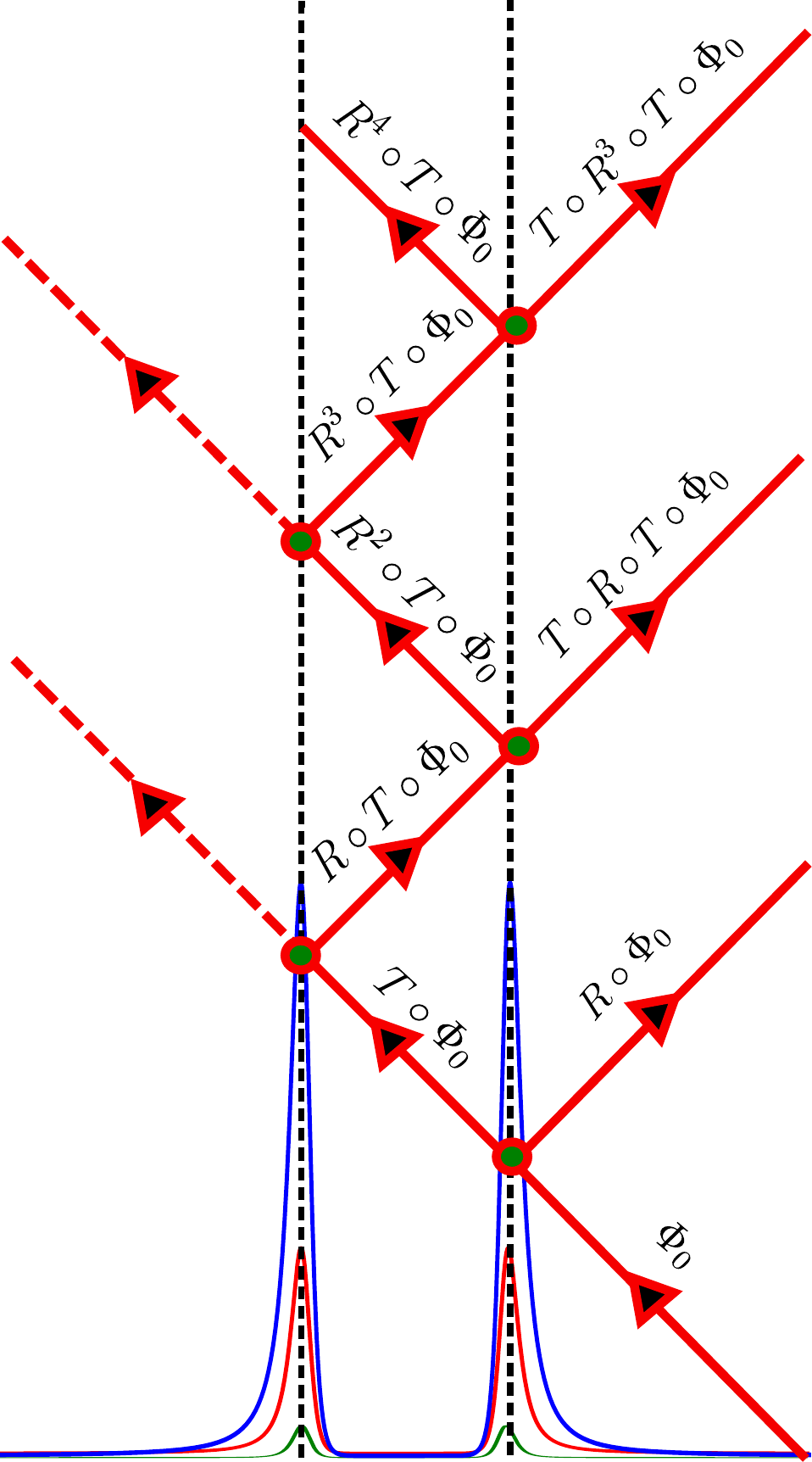}
\caption{\label{fig:Geom_optics} Schematic reconstruction of the outgoing solution by a sequence of reflections and transmissions inside the potential cavity.}
\end{figure}

\begin{figure*}
\centering
\subfigure{
\includegraphics[width=.4\textwidth]{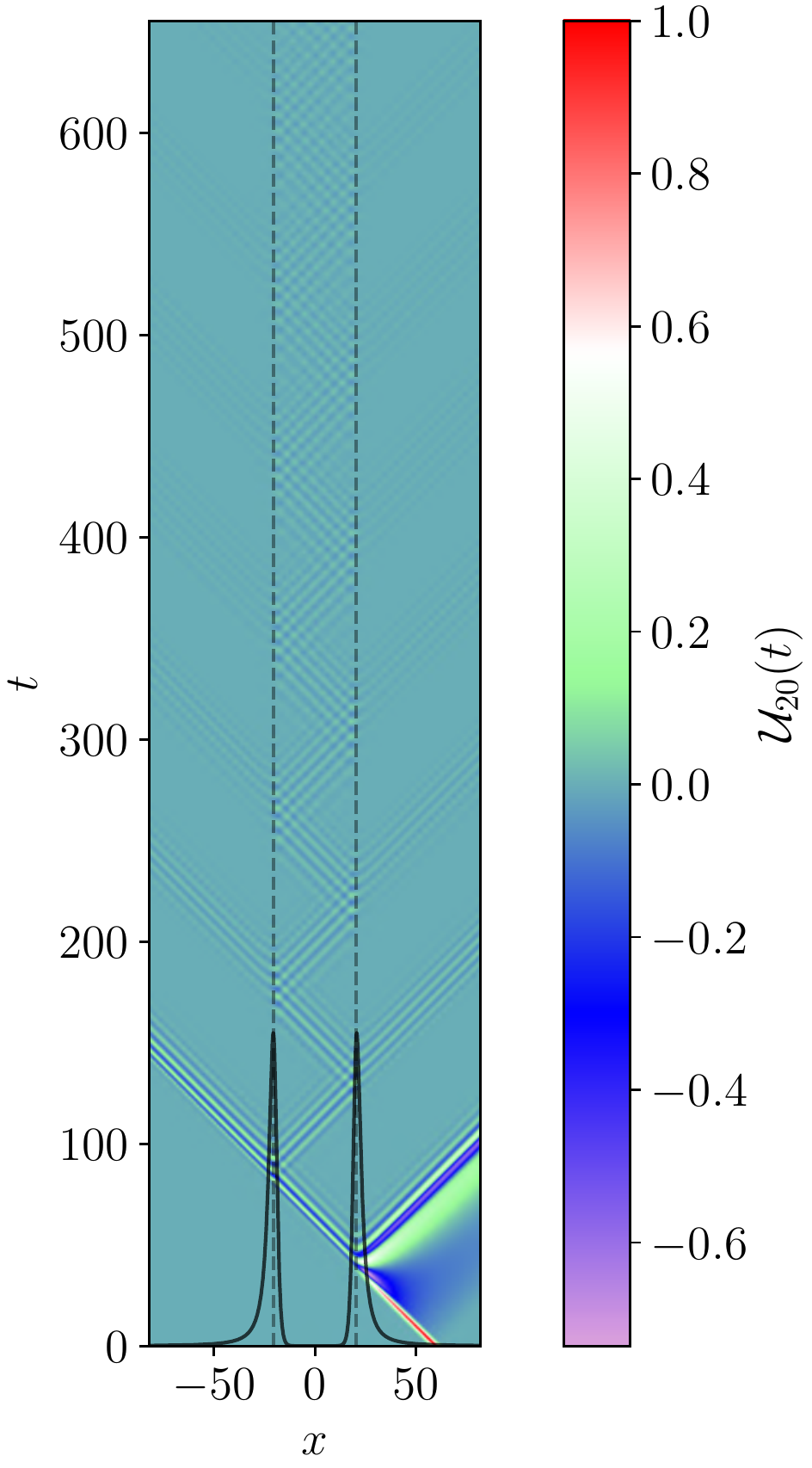}} \,
\subfigure{
\includegraphics[width=.4\textwidth]{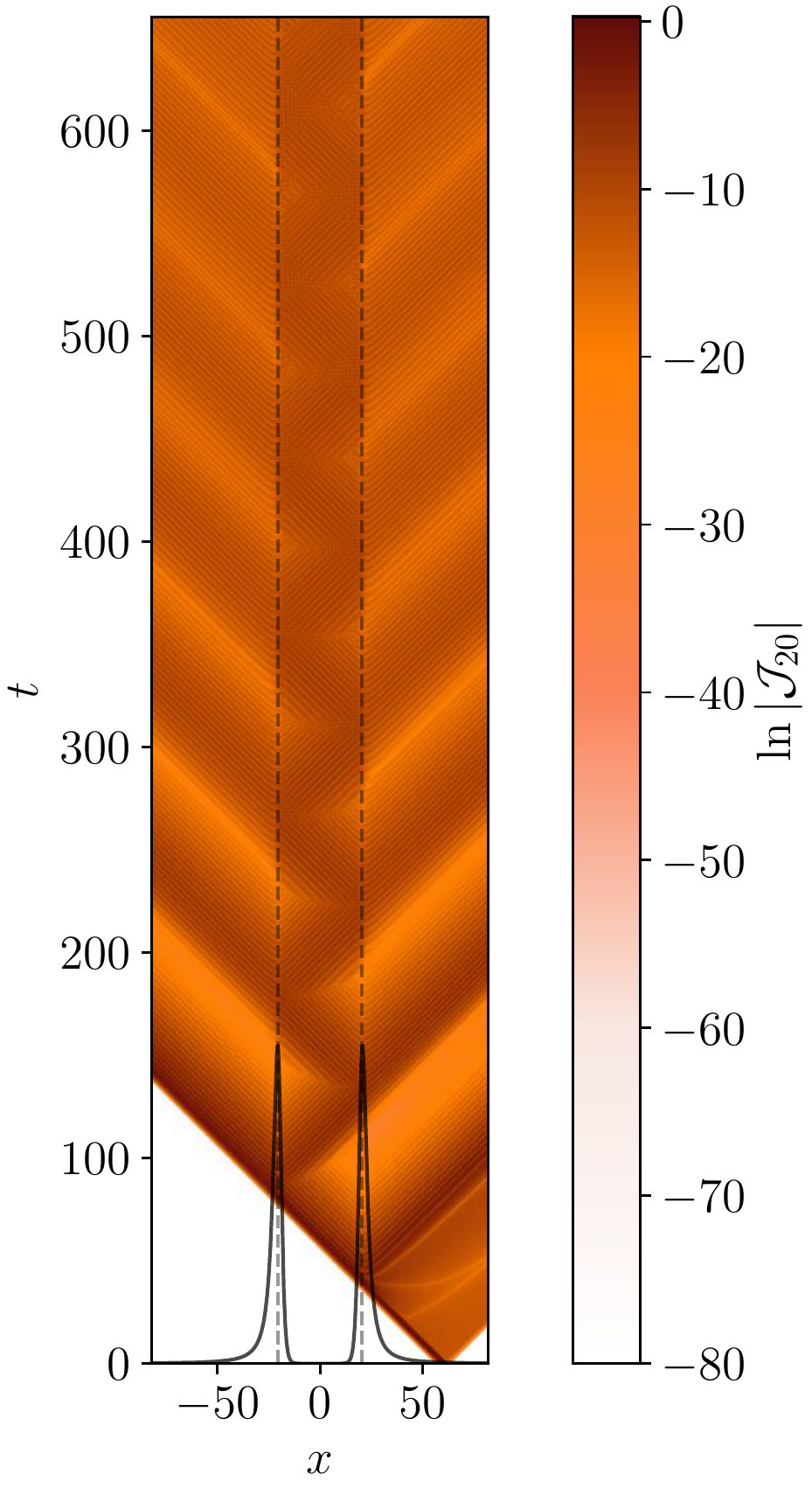}}
\caption{\label{fig:sol_n_flux_wh} Left panel: Evolution of the quadrupole mode $\mathcal{U}_{20}$ for the effective potential in Fig.~\ref{fig:potential_wh}. We can notice the sequence of reflections and transmissions inside the cavity is very similar to the scheme depicted in Fig.~\ref{fig:Geom_optics}. Right panel: Evolution of the radial flux $\mathcal{J}_{20}\equiv \Delta\mathcal{U}_{20,x} \Delta\mathcal{U}_{20,t}$, the only incident source comes from the collision of the pulse, which dissipates very slowly to the exterior every time the internal reflections hit the walls of the potential cavity. QNM are sourced by this process.} 
\end{figure*}

\begin{figure*}
\centering
\subfigure{
\includegraphics[width=.45\textwidth]{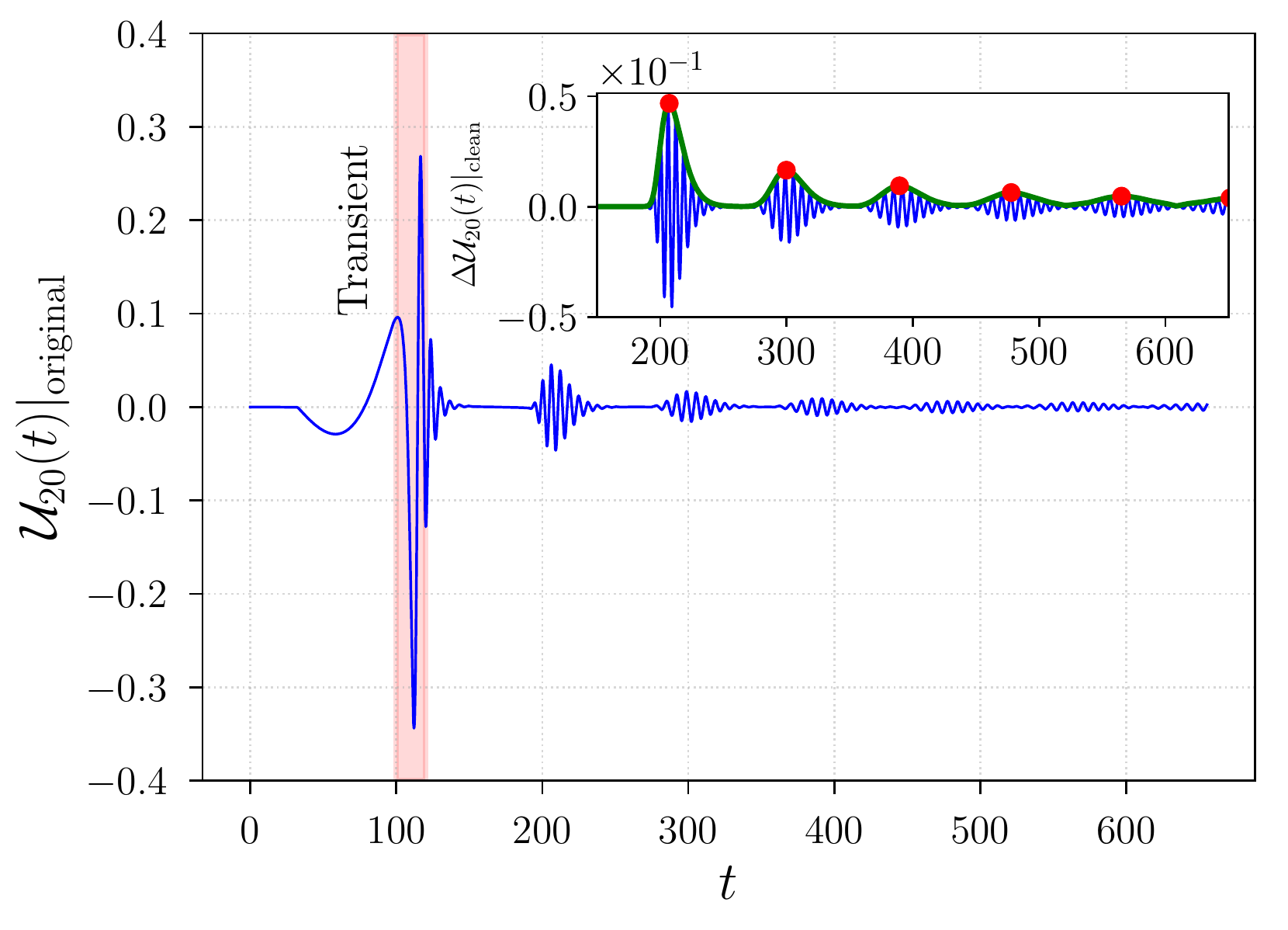}} \,
\subfigure{
\includegraphics[width=.45\textwidth]{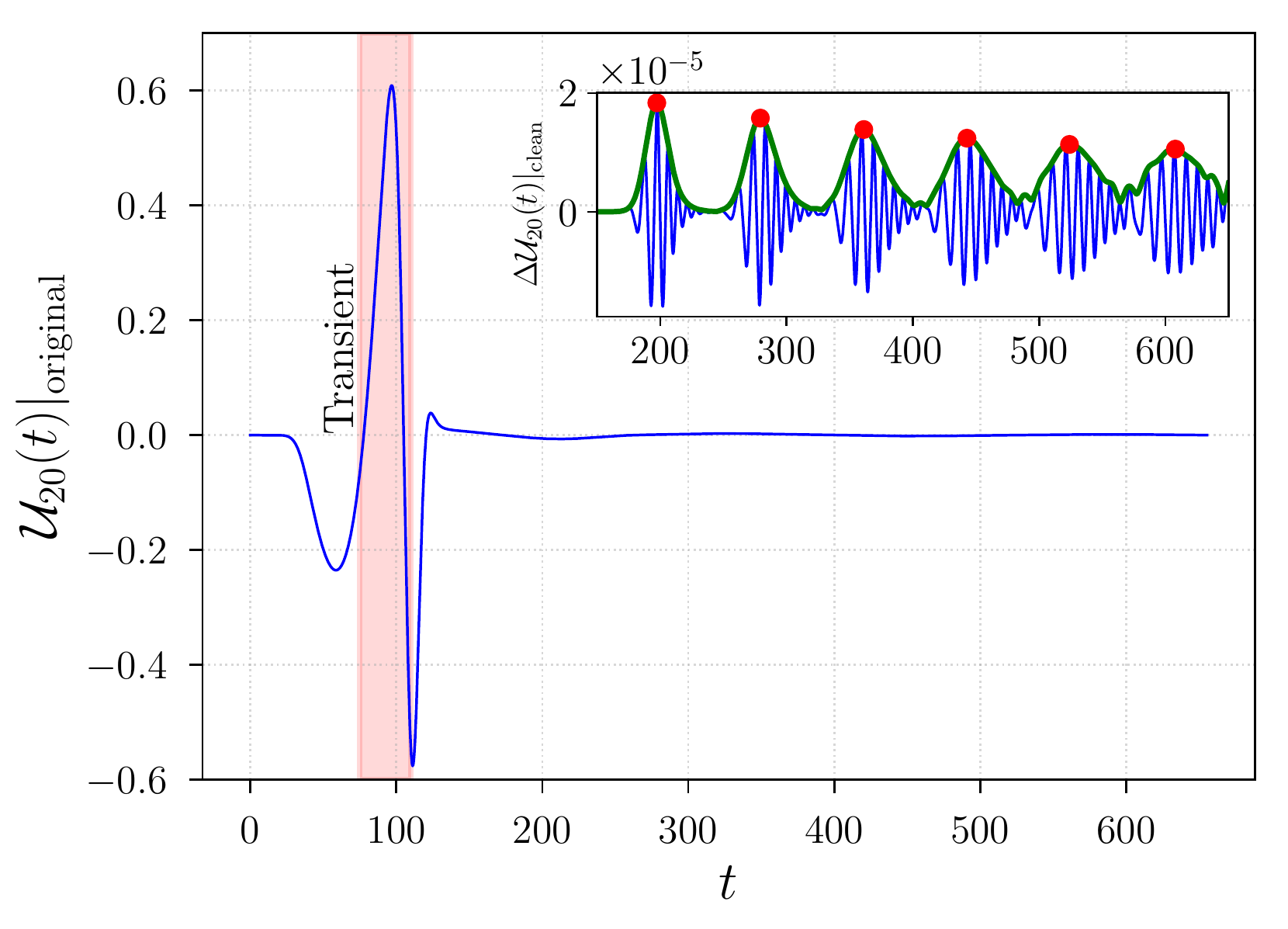}}
\caption{\label{fig:asympt_two_sigmas} Left panel: Asymptotic evolution of the quadrupole mode $\mathcal{U}_{20}$ for $\sigma=0.6495r_g$. Echoes are plotted in the upper corner of the plot, showing them along with their Hilbert envelope (the curves in green) and maximum amplitudes (in red). Right panel: Asymptotic evolution of the quadrupole mode $\mathcal{U}_{20}$ for $\sigma=5.196r_g$. In contrast with the left panel, echoes are four orders of magnitude smaller than the transient. The amplitude of the transient in the right figure has decreased with respect to the one in the left panel.} 
\end{figure*}

which is plotted in Fig.~\ref{fig:potential_wh} and coincides with the shape of the potential calculated in \citep{Cardoso:2016oxy}. Strictly speaking, we refer to $r\left(|x|\right)$ as the same inverse of the function mentioned in \eqref{eq:tortoise2} now evaluated at $|x|-r_0$. As we can see in Fig.~\ref{fig:potential_wh}, the new effective potential is merely a reflection of the potential barrier in Fig.~\ref{fig:Potential_BH} about the ordinate axis; thus, it is sensible to identify this system as a potential cavity built from two potential barriers with the reflection and transmission coefficients depicted in Fig.~\ref{fig:RT_scalar}. Furthermore, let us assume that an arbitrary incident pulse $\Phi_0$ propagates towards a cavity constituted by two different potential walls with reflectivities $R_{\mathrm{right}}(\omega)$, $R_{\mathrm{left}}(\omega)$ and transmissivities $T_{\mathrm{right}}(\omega)$, $T_{\mathrm{left}}(\omega)$. It is, therefore, reasonable to approximate the spectrum of the asymptotic solution by a simple geometrical series of reflections and transmissions inside the cavity acting on the incident pulse, as shown in Fig.~\ref{fig:Geom_optics} 

\begin{eqnarray}
&\Phi^{\mathrm{wh}}(\omega)|_{x\rightarrow+\infty}=\bigg[T_{\mathrm{right}}\circ\displaystyle{\sum_{i=0}}\left(R_{\mathrm{left}}^{i+1}\circ R_{\mathrm{right}}^{i}\right) \circ T_{\mathrm{right}}\nonumber\\
&+R_{\mathrm{right}}\bigg]\circ\Phi_0(\omega),
\label{eq:gen_reconst}
\end{eqnarray}
this expansion is in essence similar to the idea presented in \citep{Correia:2018apm}. In the simplified setup suggested in Fig.~\ref{fig:Geom_optics} for the Morris-Thorne wormhole -- studied throughout this paper -- we require two identical potential barriers following $R_{\mathrm{right}}(\omega)=R_{\mathrm{left}}(\omega)=R(\omega)$ and $T_{\mathrm{right}}(\omega)=T_{\mathrm{left}}(\omega)=T(\omega)$, which reduce the latter expression to
 \begin{equation}
\Phi^{\mathrm{wh}}(\omega)|_{x\rightarrow+\infty}=\left[R+T\circ\sum_{i=0}R^{2i+1}\circ T\right]\circ\Phi_0(\omega).
\label{eq:reconst}
\end{equation}
From \eqref{eq:gen_reconst}, it is not difficult to see consider the hypothetic case of a perfectly reflective wall $(R_{\mathrm{left}}(\omega)=1)$ replacing the left potential barrier in Figs.~\ref{fig:potential_wh} and \ref{fig:Geom_optics}, the reconstructed pulse is instead given by 
\begin{equation}
\Phi^{\mathrm{f}}(\omega)|_{x\rightarrow+\infty}=\left[R+T\circ\sum_{i=0}R^i\circ T\right]\circ\Phi_0(\omega),
\label{eq:reconst_firewall}
\end{equation}
corresponding to the case of a firewall. Nevertheless, we will not cover the features of the firewall solution in what remains of this paper. 

We now solve the scattering problem exactly for the quadrupole mode $\mathcal{U}_{20}(t,x)$. Using the same initial conditions as in \eqref{eq:scalar_init_cond} for $\sigma=0.9185r_g$ as the width of the incident Gaussian pulse, we find the time-dependent solution of the quadrupole mode $\mathcal{U}_{20}$ (left panel) and its radial flux $\mathcal{J}_{20}\equiv \Delta\mathcal{U}_{20,x} \Delta \mathcal{U}_{20,t}$ (right panel) in Fig.~\ref{fig:sol_n_flux_wh}. It is interesting to notice in the evolution plot (on the left) that the signal forms an interference pattern at very late times, showing that successive reflections might fill the cavity. In addition to this, even when the amplitude of the modes decreases after each collision against any of the potential walls, the spherical modes propagate for longer time throughout the spheres of maximal potential. 

The scalar flux is shown in the right panel, we observe that the only source of scalar radiation comes from the first collision of the Gaussian wavepacket against the barrier in the right hand side (the ingoing flux is colored in black at the bottom of the contour plot). A sequence of reflections occurs within the potential barriers, which decay in intensity with time as the cavity leaks energy to the exterior.

Amplitudes of the outgoing signals depend on the width of the incident Gaussian wavepacket, this is visible in Fig.~\ref{fig:asympt_two_sigmas} where we plot the asymptotic solutions for two different ingoing wavepackets: one with $\sigma=0.6495r_g$ in the left panel and a second one with $\sigma=5.196r_g$ in the right. After subtracting the outgoing solution for a black hole, the presence of a train of wavelets, colloquially known as echoes, is very clear. For large values of $\sigma$, the signal is not featureless after the transient, as we can observe in the right panel of the same figure. Therefore, subtracting the outgoing pulse (i.e., the case in which there is only one potential wall) obtained from the black hole is a convenient way to clean the signal from back reflections due to the ``tails'' of the potential barrier. The necessity of this procedure is more evident in the case depicted in the right panel of Fig.~\ref{fig:asympt_two_sigmas}, where the amplitudes of the echoes are four orders of magnitude smaller than the transient. In both panels, we plot the variable $\Delta\mathcal{U}_{20}|_{\mathrm{clean}}\equiv\mathcal{U}_{20}|_{\mathrm{original}}-\mathcal{U}^{\mathrm{bh}}_{20}$ in the right upper corner of the figures to represent the echoes and their net amplitude after removing backscattering effects. Notice that, in the upper corner of both figures, the amplitude of the echoes does not decrease exponentially with time due to the successive reflections inside the cavity.
 
As shown in subsection \ref{subsec:bh}, the curves of reflectivity and transmissivity determines which frequencies stay in the cavity: most of the power of an incident pulse with large $\sigma$ is in the low frequency domain, and therefore it will be reflected. The cavity is transparent to high frequency signals, which are dominant in the pulses with small $\sigma$. In either of these extremal scenarios, QNMs cannot be sourced by internal reflections and thus, the amplitude of the echoes is not large in general. Furthermore, the steepness of the transition near the overlap point $R^2(\omega)=T^2(\omega)=0.5$ regulates the abundance of frequencies in the spectrum of outgoing signals.\\
 
\begin{figure}[t!]
\centering
\includegraphics[width=.45\textwidth]{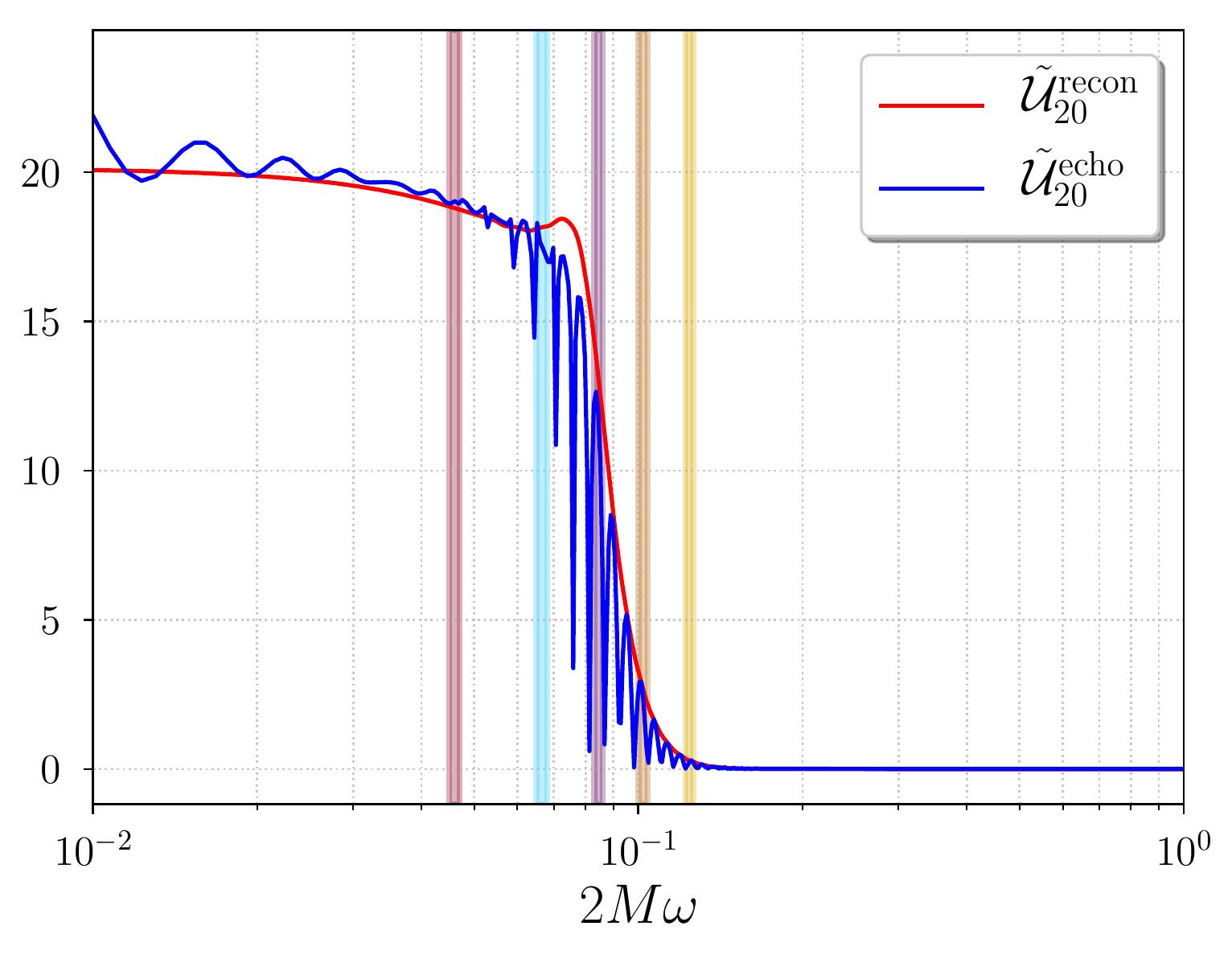}
\caption{\label{fig:rec_scalar} Comparing the reconstruction of the spectrum by the geometrical optics approximation in \eqref{eq:reconst} (in red) with the Fourier transform of the full asymptotic solution $\tilde{\mathcal{U}}_{20}^{\mathrm{echo}}$ (in blue). The color stripes indicate a few of the first QNM frequency peaks corresponding to $\omega_3=0.251M^{-1}$ in golden rod, $\omega_4=0.207M^{-1}$ in peru, $\omega_5=0.169M^{-1}$ in plum, $\omega_6=0.133M^{-1}$ in light blue and $\omega_7=0.092M^{-1}$ in rosy brown, these are the real parts of the QNM frequencies for $n=3,4,5,6~\&~7$ calculated in \citep{PhysRevD.46.4179}.}
\end{figure} 

As a next step, we proceed to reconstruct the asymptotic spectrum by following the geometrical optics relation in \eqref{eq:reconst} considering the reflectivity and transmissivity operators defined in \eqref{eq:ref_and_trans}. Henceforth, the outcome should be compared with the spectral content of the asymptotic wave solutions of \eqref{eq:wave_scalar_wh}. We calculate the Fourier transforms of both the Gaussian incident pulse $\Phi_0(\omega)=\mathcal{F}[\exp\left((x-x_0)^2/2\sigma^2\right)]$ for $\sigma=0.6495r_g$ and $x_0=60.0r_g$, and the asymptotic solution $\tilde{\mathcal{U}}_{20}^{\mathrm{echo}}\equiv\mathcal{F}[\mathcal{U}_{20}(t,+\infty)]$ including the echoes.\\

\begin{figure*}
\centering
\subfigure{
\includegraphics[width=.45\textwidth]{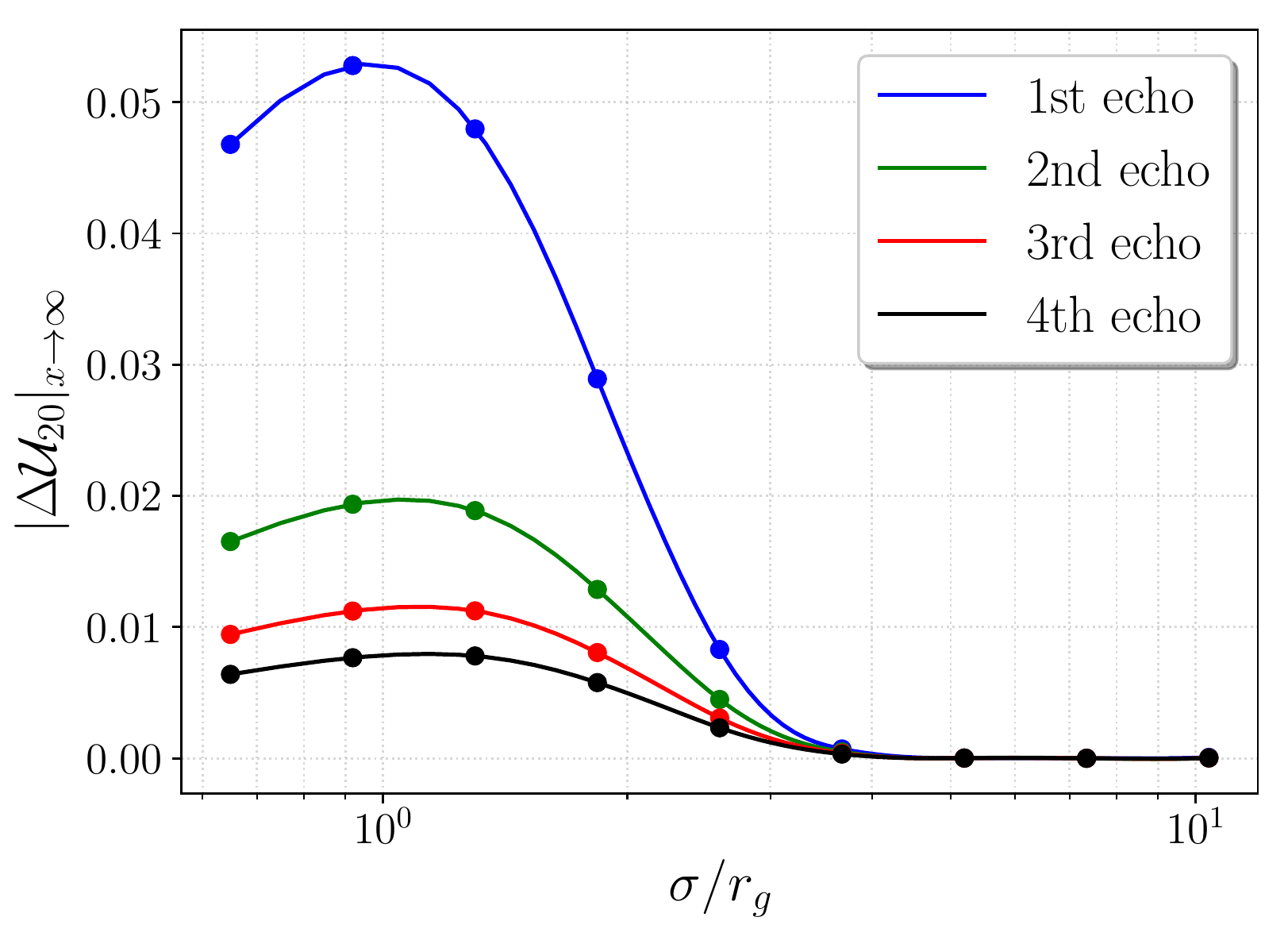}} \,
\subfigure{
\includegraphics[width=.45\textwidth]{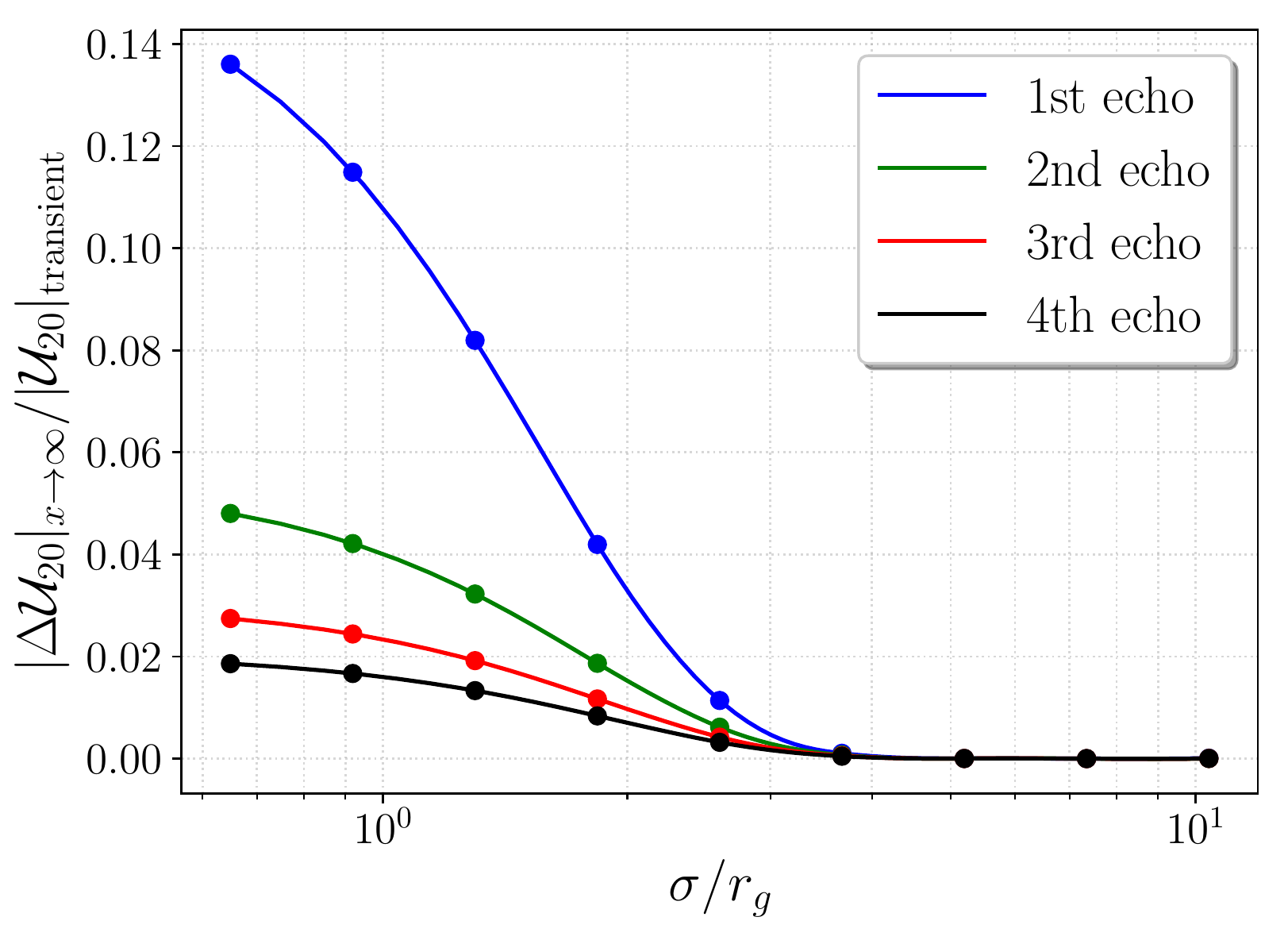}}
\caption{\label{fig:echo_sigma} Left panel: Amplitudes of the first four echoes as a function of $\sigma$ for $\ell=2$, we observe the presence of a maximum amplitude of all the echoes at roughly $\sigma\approx r_g$, which is not incompatible with the de DeWitt approximation. Right panel: Relative amplitude of the first four echoes compared to the amplitude of the transient. The points represent the simulated wormhole/black hole pairs used in our analysis.} 
\end{figure*}

After applying the reconstruction expression in \eqref{eq:reconst} up to $i=0$ (in red), and comparing the outcome with $\tilde{\mathcal{U}}_{20}^{\mathrm{echo}}$, we show the reconstructed spectrum and $\tilde{\mathcal{U}}_{20}^{\mathrm{echo}}$ (in blue) in Fig.~\ref{fig:rec_scalar}. The low frequency oscillation peaks in the blue spectrum correspond to the finite size of the simulation box and should be ignored. Notice that the spectrum reconstructed employing the geometrical optics approximation gives the overall shape of the spectrum with decent precision but not the QNM frequency peaks, these appear in the same frequency interval where the reflectivity and transmissivity curves intersect in Fig.~\ref{fig:RT_scalar}. It is, therefore, reasonable to talk about a ``sweet spot'' in the frequency domain where the cavity maximizes the amplitude of the echoes. Intuitively, after observing the results in Fig.~\ref{fig:asympt_two_sigmas}, it is possible to identify a similar ``sweet spot'' in the parameter space space for the widths of the incident Gaussian wavepackets, considering this is a one parameter problem. However, it is also necessary to not only compare the amplitude of each individual echo with $\sigma$, but also the ratio between the amplitude of the echo with the amplitude of the transient for each value of the width, which is relevant since we are finding the relative intensity of the echoes compared to the strongest outgoing signal. To do so, we proceed as follows: we setup a logarithmic 1D grid of widths centered at $\sigma_{\mathrm{DW}}=\sqrt{27}r_g/2\ell$ --  the width of maximum transmissivity according to the DeWitt approximation \citep{DeWitt:1975ys} -- and spaced by increments of $\sqrt{2}\sigma_{\mathrm{DW}}$. Then, we find the amplitudes of the first four echoes for every width of the incident pulse without changing the cavity. In order to determine the amplitudes of the echoes, we need to subtract the reflection coming from the scattering of a single barrier (i.e., the black hole case) by using the variable $\Delta\mathcal{U}_{20}|_{\mathrm{clean}}\equiv\mathcal{U}_{20}|_{\mathrm{original}}-\mathcal{U}^{\mathrm{bh}}_{20}$. This requires a non-trivial computational effort since each scattering scenario needs to be solved twice (one for the wormhole and one for the black hole) in order to clean up the signal and obtain a clear view of the echoes. Once the signal is refined, we find the continuous envelope of the asymptotic solution by calculating its Hilbert transform \citep{1449626}, represented as the green curves in the upper corners of Fig.~\ref{fig:asympt_two_sigmas}, and therefore, the amplitudes of the echoes are the local maxima of these envelopes (the red dots in the same figure). 

Our results can be found in Fig.~\ref{fig:echo_sigma}, where we notice in the left panel the presence of a well-defined maximal amplitude of the first four echoes. This is consistent with the idea of a range of frequencies/widths that maximize the amplitudes of the echoes, as shown in the analysis of the reflection and transmission coefficients: the cavity is transparent for small widths of the ingoing Gaussian (large frequencies) and is reflective for wide incident pulses. In the right panel, we observe the growth of the relative amplitude as the widths become smaller. Such a fact only means that the reduction of the transient is faster than the reduction of the echoes as the frequencies grow, as the cavity becomes more transparent the ingoing pulses get transmitted more efficiently. Consequently, this analysis is useful to get a basic understanding of the frequency/width scales in which echoes could be observable.  

\begin{figure*}
\centering
\subfigure{
\includegraphics[width=.45\textwidth]{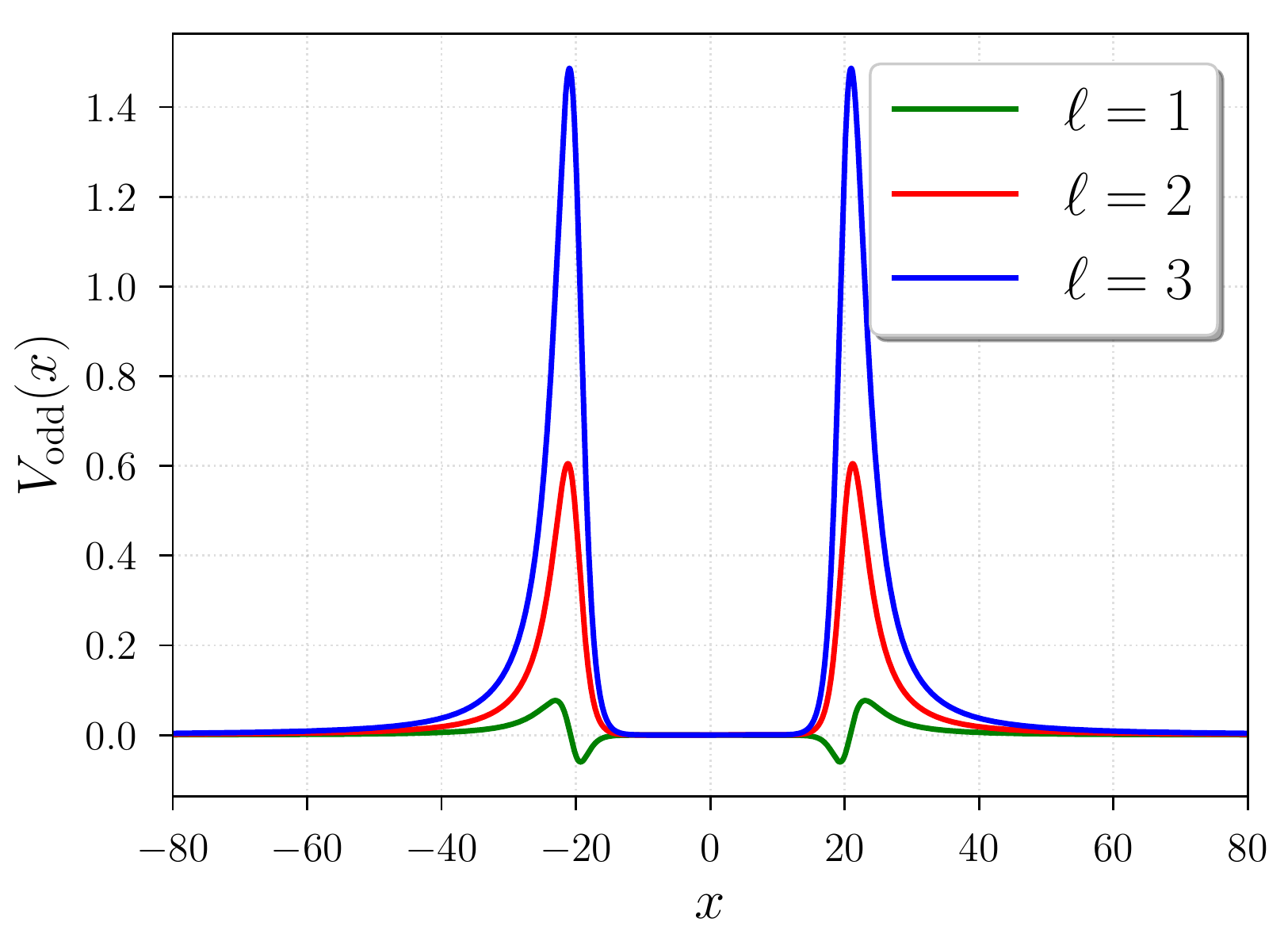}} \,
\subfigure{
\includegraphics[width=.45\textwidth]{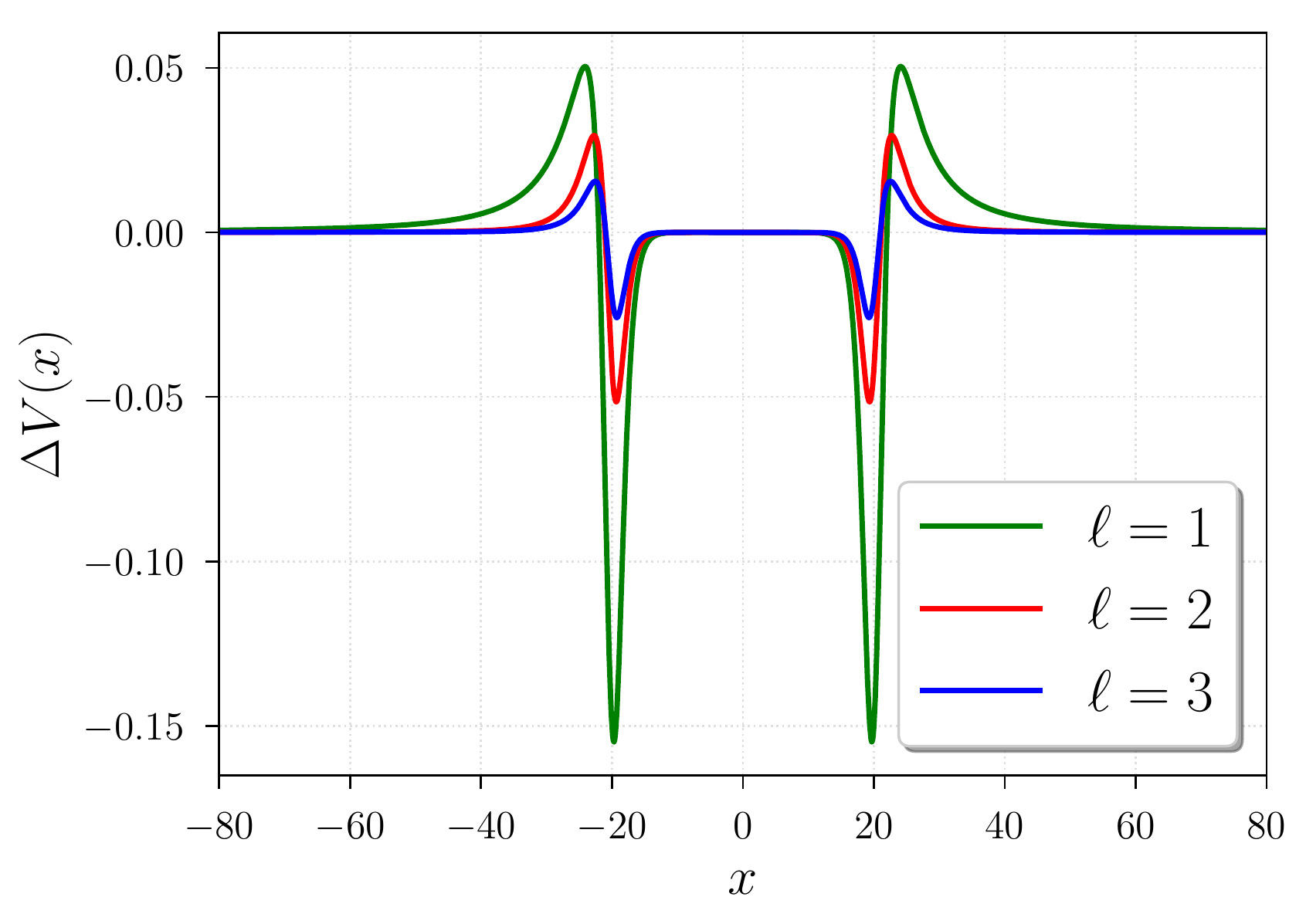}} 
\caption{\label{fig:evenodd_potentials} Left panel: Plot of the Regge-Wheeler potential ($V_{\mathrm{odd}}\equiv[1-r_g/r(|x|)]\mathcal{V}_{\mathrm{odd}}$) as a function of the tortoise coordinate $x$. Right panel: Difference between the Regge-Wheeler and Zerilli effective potentials $\Delta V=V_{\mathrm{odd}}-V_{\mathrm{even}}$. The difference of the two potentials is below the order of 1\% for ($\ell\geq 2$).} 
\end{figure*}

\section{Scattering of tensor wavepackets}\label{sec:tensor} 
Following the perspective of the even/odd parity decomposition for tensor perturbations of a spherically symmetric spacetime \citep{Regge:1957td, PhysRevD.2.2141, PhysRevD.5.2419}, we implement all the techniques used in section \ref{sec:scalar} and extend our analysis to study the scattering of a test Gaussian wavepacket of tensor perturbations. From now on, we will follow the conventions in \citep{Martel:2005ir}, including the choice of the Regge-Wheeler gauge. 
The dynamics of the wave scattering problem is given by two equations of motion of the form
\begin{equation}
\left(\tilde{\Box}-\mathcal{V}_{\mathrm{eff}}\right)\Psi_{\ell m}=0,
\label{eq:tensor_eq_mov}
\end{equation} 
here $\tilde{\Box}\equiv g^{ab}\nabla_a\nabla_b$ is the 2D d'Alembertian operator in the usual $(a,b)\rightarrow(t,r)$ Schwarzschild coordinates. $\mathcal{V}_{\mathrm{eff}}$ corresponds to one of two possible potentials, the Regge-wheeler (odd) potential, $\mathcal{V}_{\mathrm{odd}}$ 
\begin{equation}
\mathcal{V}_{\mathrm{odd}}(r)= \frac{\ell(\ell+1)}{r^2}-\frac{3r_g}{r^3},
\label{eq:V_odd}
\end{equation} 
or the Zerilli (even) potential, $\mathcal{V}_{\mathrm{even}}$ 
\begin{equation}
\mathcal{V}_{\mathrm{even}}(r)=\displaystyle{\frac{1}{\Lambda^2}\bigg[\mu^2\bigg(\frac{\mu+2}{r^2}+\frac{3r_g}{r^3}\bigg)+\frac{9r_g^2}{r^4}\left(\mu+\frac{r_g}{r}\right)\bigg]},
\label{eq:V_even}
\end{equation} 
where $\mu\equiv(\ell-1)(\ell+2)$ and $\Lambda\equiv\mu+3r_g/r$. All the source terms proportional to the stress energy tensor and its contractions appearing in the right hand side of \eqref{eq:tensor_eq_mov} in \citep{Martel:2005ir} are not considered for the scattering problem. The introduction of tortoise coordinates $(t,x)$ is also very convenient and works in exactly the same way as in \eqref{eq:tortoise} and \eqref{eq:tortoise2}, in these coordinates the waveform of the two equations of motion -- one for the odd parity modes and another for the even -- is given by 
\begin{equation}
\left[-\partial_t^2+\partial_x^2-V_{\mathrm{eff}}(x)\right]\Psi_{\ell m}(x,t) = 0,
\label{eq:wave_tensor_wh}
\end{equation}
where $V_{\mathrm{even}}\equiv[1-r_g/r(|x|)]\mathcal{V}_{\mathrm{even}}$ and $V_{\mathrm{odd}}\equiv[1-r_g/r(|x|)]\mathcal{V}_{\mathrm{odd}}$. Recalling the procedure followed in subsection \ref{subsec:wh}, our setup already includes the effective potentials for the Morris-Thorne wormholes, obtained by reflection of the potential barriers about the ordinate axis. The effective potentials are plotted in the left panel of Fig.~\ref{fig:evenodd_potentials}, which is very similar to the one in the scalar scattering. In the right panel it is possible to notice that the difference between the potentials is only substantial at $(\ell\leq1)$. Both the even and odd solutions of \eqref{eq:tensor_eq_mov} and \eqref{eq:wave_tensor_wh} are already spherical modes used to find the two asymptotic polarizations of the tensor fluctuations propagating in a flat background, such as the term in the diagonal, $h_+$ 
\begin{align}
\displaystyle{h_+=\frac{1}{r(|x|)}\sum_{\ell,m}\bigg\{\Psi_{\ell m}^{\mathrm{even}}\left[\partial^2_{\theta}+\frac{1}{2}\ell(\ell+1)\right]Y_{\ell m}(\theta,\phi)}\nonumber\\
\displaystyle{-\Psi_{\ell m}^{\mathrm{odd}}\frac{i m}{\sin\theta}\left[\partial_{\theta}-\frac{\cos\theta}{\sin\theta}\right]Y_{\ell m}(\theta,\phi)\bigg\}},
\label{eq:hplus}
\end{align} 
and the off-diagonal, $h_{\times}$ 
\begin{align}
\displaystyle{h_{\times}=\frac{1}{r(|x|)}\sum_{\ell,m}\bigg\{\Psi_{\ell m}^{\mathrm{odd}}\left[\partial^2_{\theta}+\frac{1}{2}\ell(\ell+1)\right]Y_{\ell m}(\theta,\phi)}\nonumber\\
\displaystyle{+\Psi_{\ell m}^{\mathrm{even}}\frac{i m}{\sin\theta}\left[\partial_{\theta}-\frac{\cos\theta}{\sin\theta}\right]Y_{\ell m}(\theta,\phi)\bigg\}}.
\label{eq:hx}
\end{align}

From these expressions, it is simple to see that the the monopole ($\ell=0$) and the dipole ($\ell=1$) terms are identically zero. Thus, the first nontrivial contributions come from the quadrupole solutions $\Psi_{20}^{\mathrm{odd}}(x,t)$ and $\Psi_{20}^{\mathrm{even}}(x,t)$, from which the differences in the odd and even potentials are small, and become even smaller for every $\ell>3$, as we can see in right panel of Fig.~\ref{fig:evenodd_potentials}. Additionally, it is reasonable to identify $h_+$ with the even mode and  $h_{\times}$ with the odd in the equatorial plane up to a constant. Hence, our analysis for the scattering dynamics and the reflected/transmitted frequencies does not require from both the even and odd solutions of \eqref{eq:wave_tensor_wh} to extend the discussions from section \ref{sec:scalar}. However, we will explain one of the consequences of the difference between the Regge-Wheeler and Zerilli potentials in subsection \ref{sec:Subsec_Polar}.

\begin{figure}[t!]
\centering
\includegraphics[width=.45\textwidth]{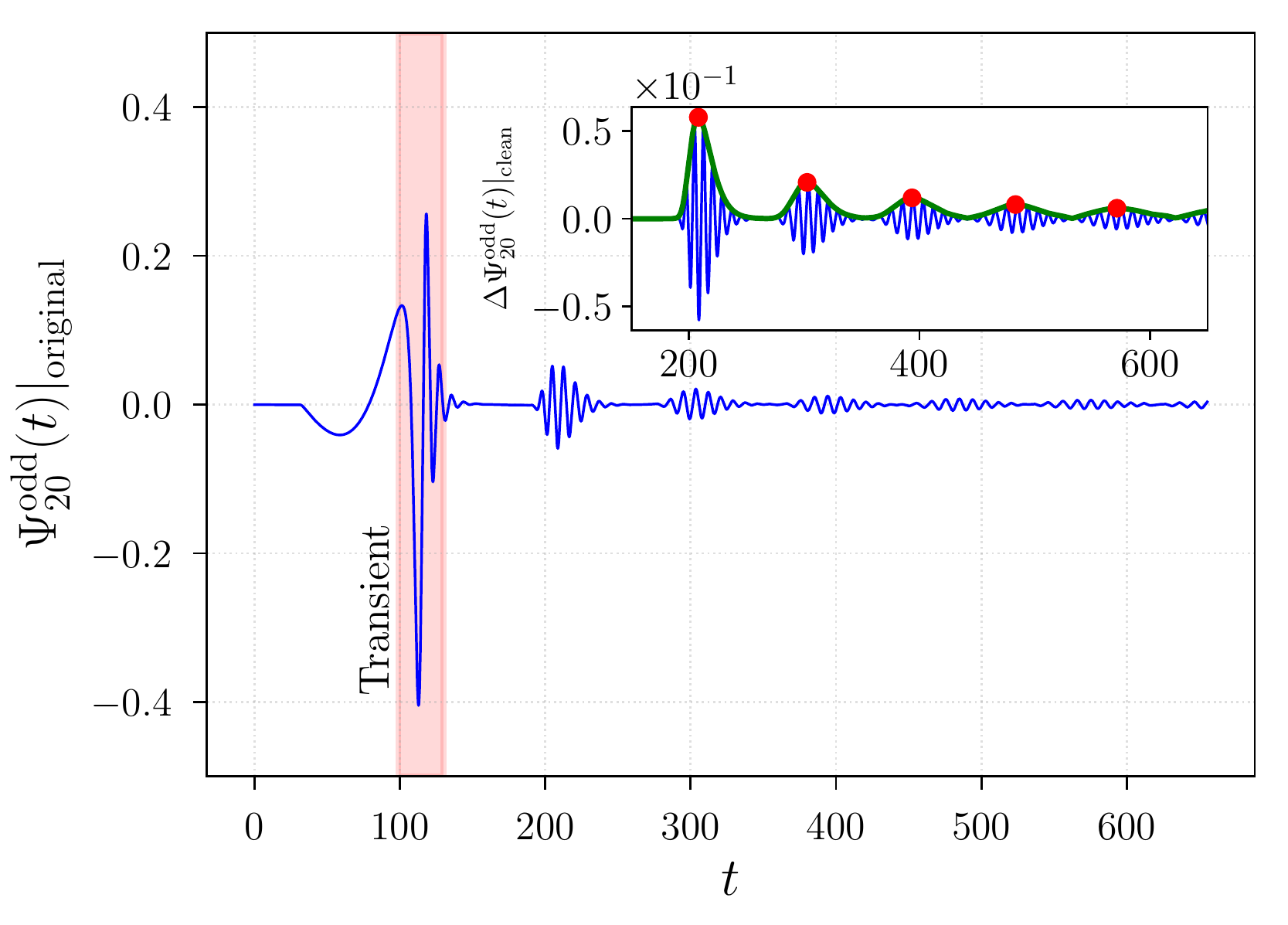}
\caption{\label{fig:odd_sigma_small} Asymptotic solution of $\Psi^{\mathrm{odd}}_{20}$ considering an incident wavepacket with $\sigma=0.9185r_g$}
\end{figure}

\begin{figure}[t!]
\centering
\includegraphics[width=.45\textwidth]{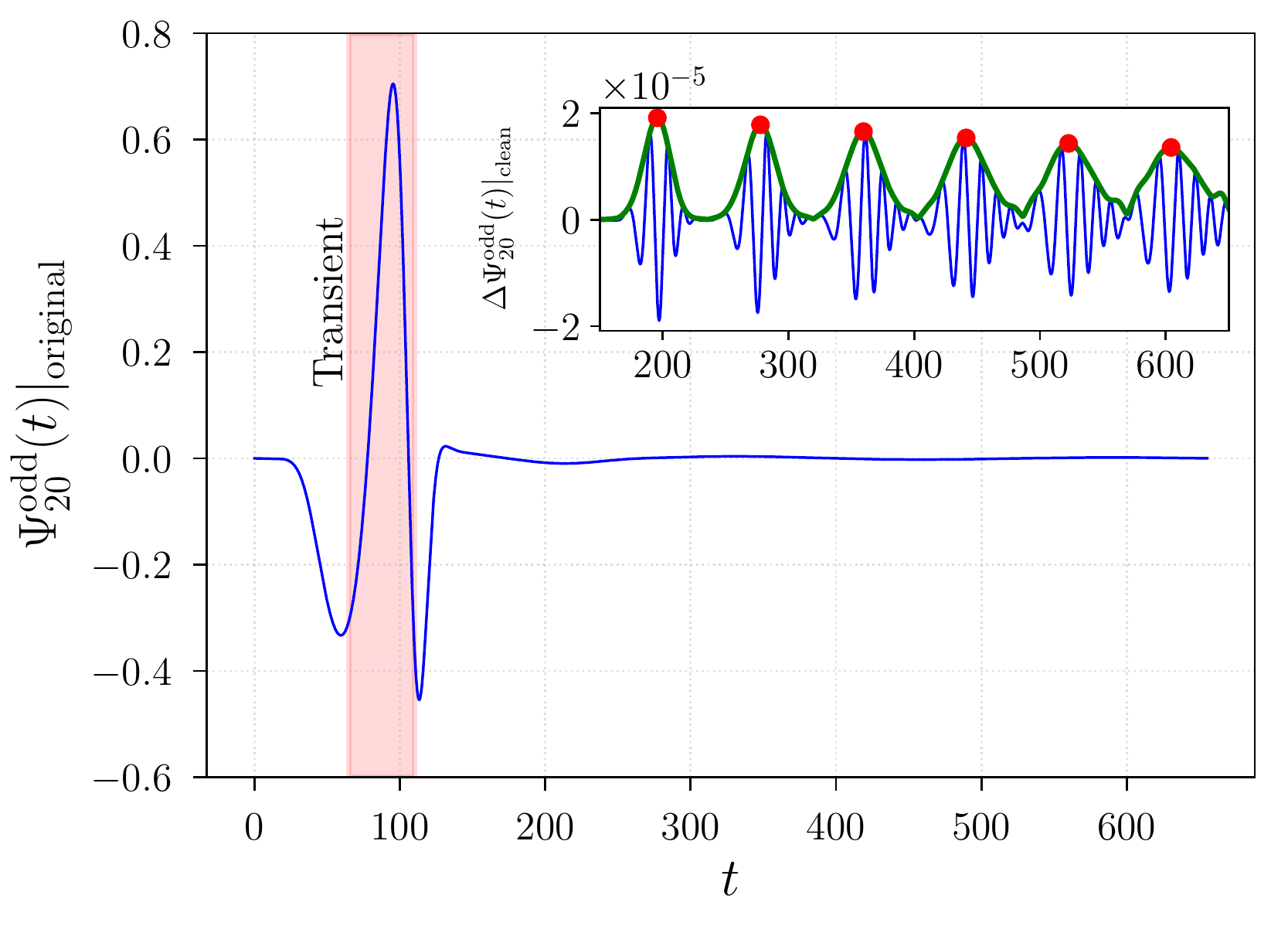}
\caption{\label{fig:odd_sigma_large} Asymptotic solution of $\Psi^{\mathrm{odd}}_{20}$ considering an incident wavepacket with $\sigma=5.196r_g$. The amplitude of the echoes is four orders of magnitude smaller than the transient.}
\end{figure}

\begin{figure}[t!]
\centering
\includegraphics[width=.4\textwidth]{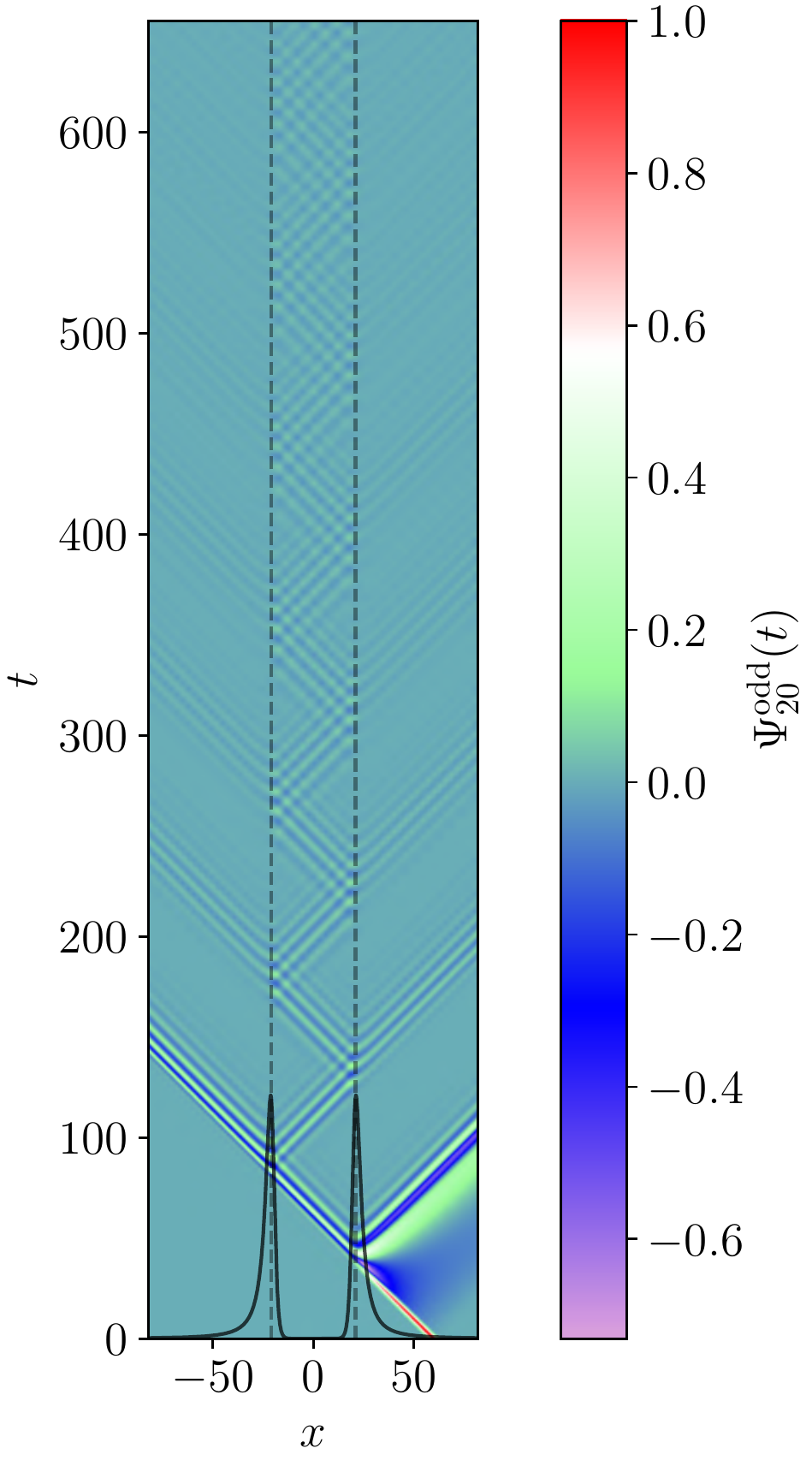}
\caption{\label{fig:even_evolve} Evolution of $\Psi^{\mathrm{odd}}$ for an ingoing Gaussian pulse with $\sigma=0.9185r_g$.}
\end{figure}

\begin{figure*}[t!]
\centering
\subfigure{
\includegraphics[width=.45\textwidth]{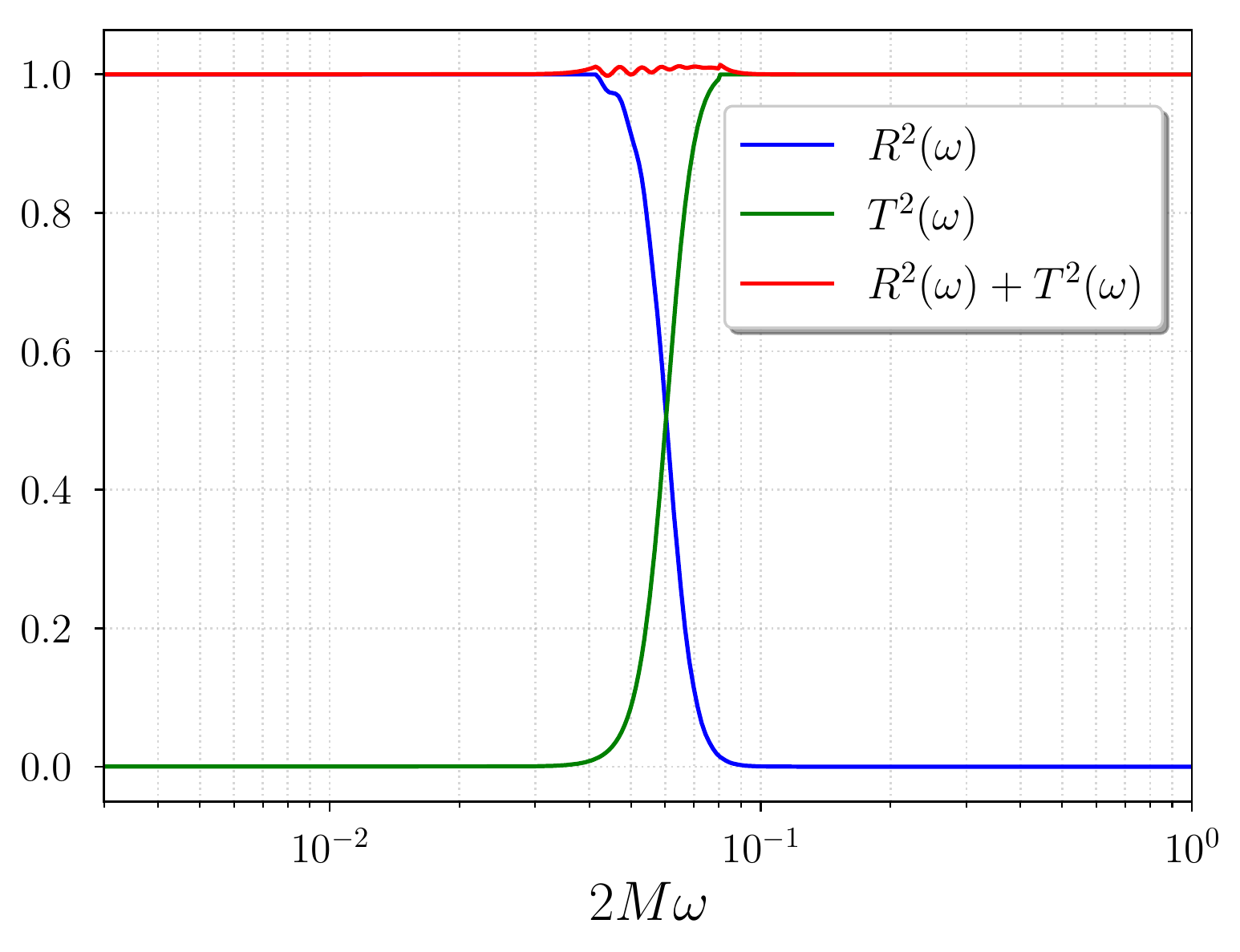}} \,
\subfigure{
\includegraphics[width=.45\textwidth]{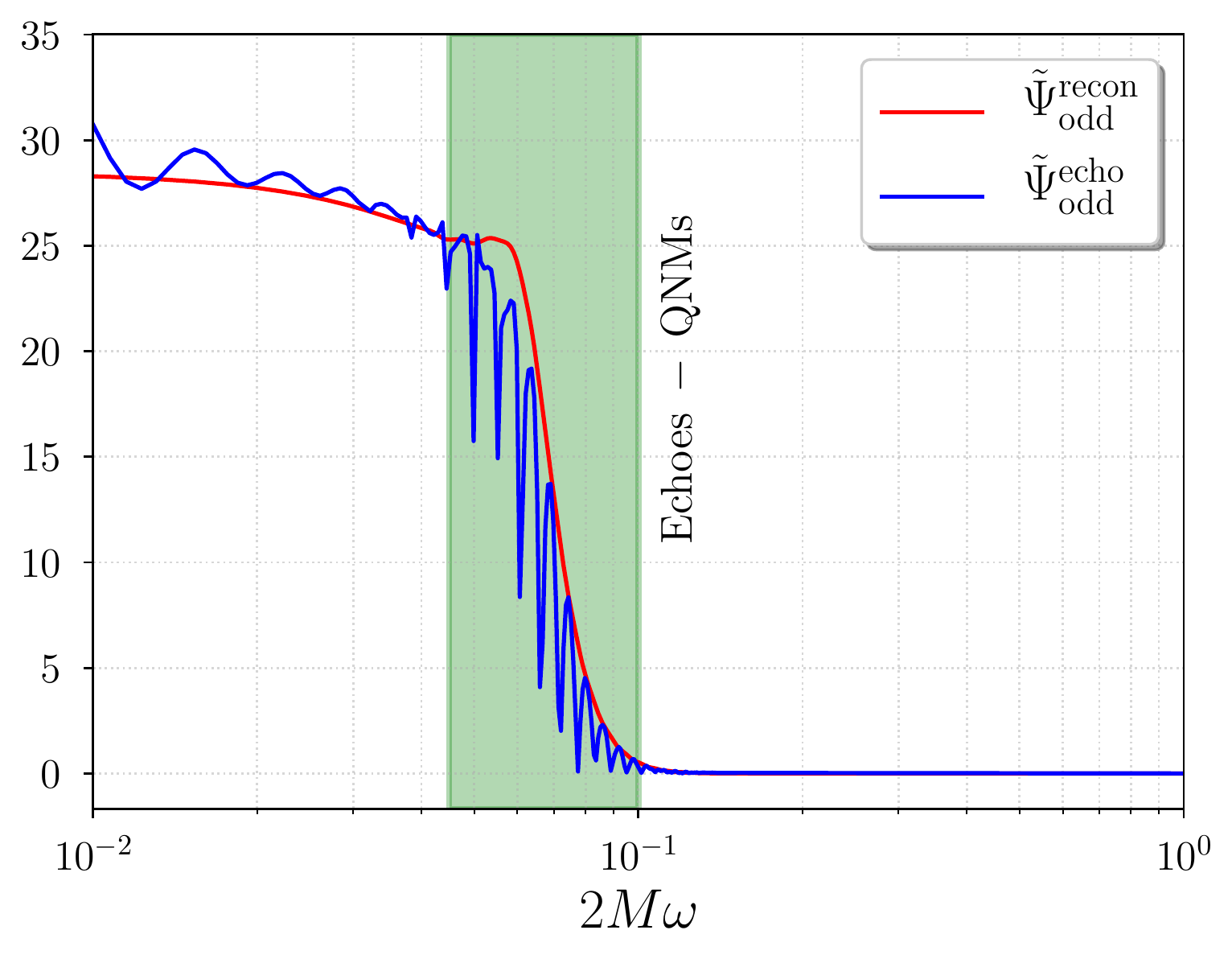}}
\caption{\label{fig:RT_reconst} Left panel: Reflection and transmission coefficients for $\sigma=0.9185r_g$ and the quadrupole $\ell,m=(2,0)$, the ``sweet spot'' in the frequency domain is located around the intersection at $R^2=T^2=0.5$. Right panel: Geometrical optics reconstruction is plotted in red, and it is compared with the Fourier transform of the asymptotic solution shown in Fig.~\ref{fig:odd_sigma_small}. Ignoring the low frequency peaks (introduced by the finite size of the simulation box), we notice that the reconstructed spectrum provides a good idea of the overall shape, but it does not reproduce the power in the frequency of the QNMs. Our results are not dramatically different for the solutions of $\Psi^{\mathrm{even}}$.} 
\end{figure*}

In analogy with the previous section, now we solve the equations of motion for the scattering process. Our setup for the initial conditions of $\Psi_{20}^{\mathrm{odd}}(x,0)$ and $\Psi_{20}^{\mathrm{even}}(x,0)$ and their time derivatives is not different from \eqref{eq:scalar_init_cond}  
\begin{align}
\displaystyle{\Psi^{\mathrm{odd}}_{20}(x,0)=\exp\left(\frac{(x-x_0)^2}{2\sigma^2}\right)},\nonumber\\
\displaystyle{\partial_t \Psi^{\mathrm{odd}}_{20}(x,t)\bigg{|}_{t=0}=\partial_x\Psi^{\mathrm{odd}}_{20}(x,0)},
\label{eq:tensor_init_cond}
\end{align}
and the same applies for $\Psi^\mathrm{even}_{20}$ and its initial time derivative. Using $\sigma=0.9185r_g$, the same initial position of the Gaussian wavepackets -- i.e., $x_0=60.0r_g$ -- and the same separation between potential walls -- i.e., $r_0=20.0r_g$ -- as before. We show the evolution of $\Psi_{20}^{\mathrm{odd}}(x,t)$ in Fig.~\ref{fig:even_evolve}, where the dispersion of the ingoing pulse is not significantly different from our results in the left panel of Fig.~\ref{fig:sol_n_flux_wh}: this is not surprising due to the similarities between the shapes of the effective potentials for scalar and tensor modes, which seem to become even more similar for higher values of $\ell$. At late times, the cavity is filled showing an interference pattern. Internal reflections make the QNMs propagate for longer in the spheres of maximum effective potential.

In Figs.~\ref{fig:odd_sigma_small} and \ref{fig:odd_sigma_large} we see the asymptotic behavior of two solutions with $\sigma=0.9185r_g$ and $\sigma=5.196r_g$, respectively. Observing that the amplitude of the echoes is not large in general, since it varies depending on the spectral content of the initial pulses, which are not the same in the case of initial Gaussian wavelets with different widths. In analogy with the scalar case, we notice in the upper corner of both figures that the amplitude of the echoes does not decay exponentially in time.
\par
\begin{figure*}
\centering
\subfigure{
\includegraphics[width=.45\textwidth]{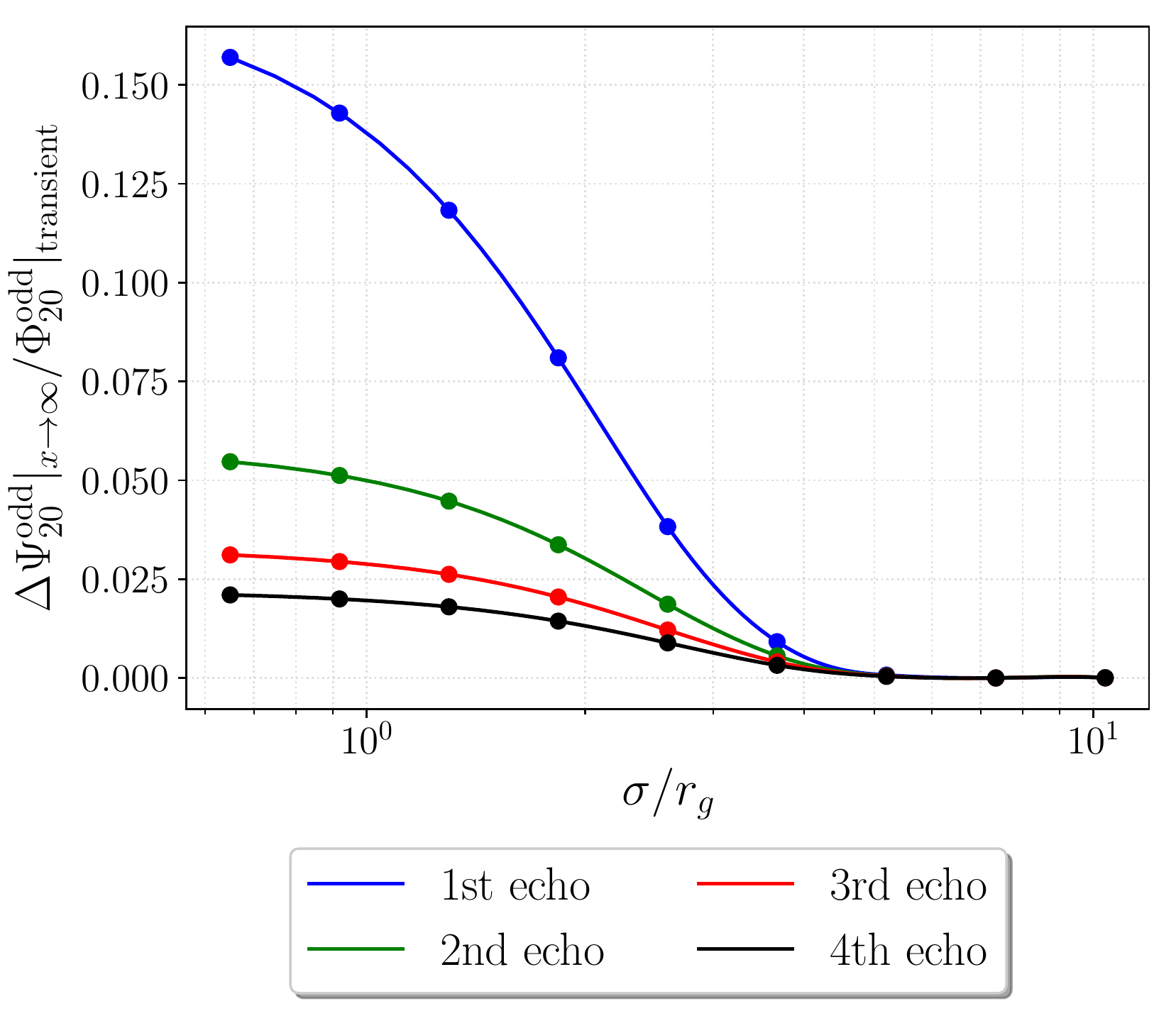}} \,
\subfigure{
\includegraphics[width=.45\textwidth]{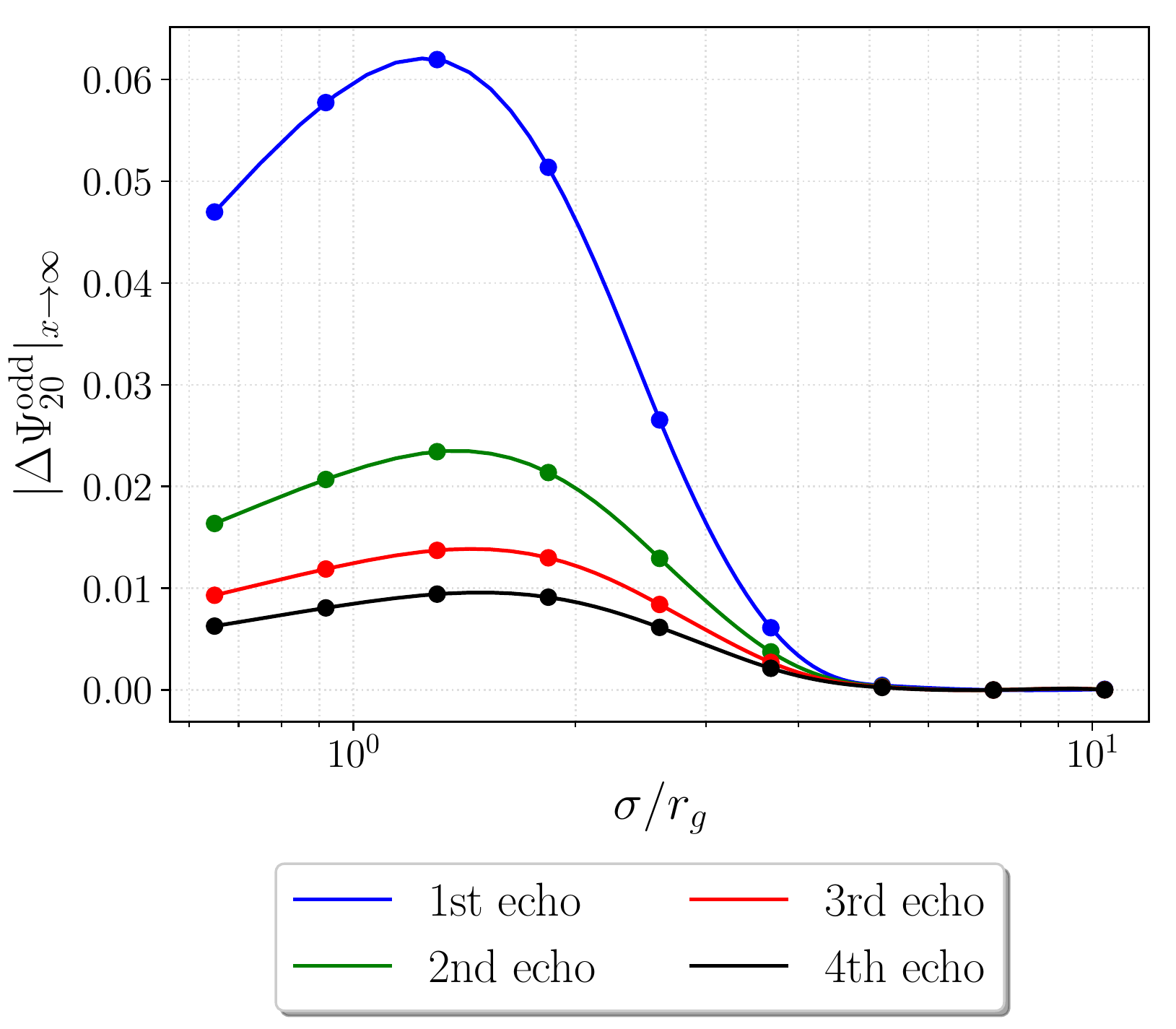}} 
\caption{\label{fig:echo_sigma_odd} Left panel: Amplitudes of the first four echoes of $\Psi^{\mathrm{odd}}_{20}$ as a function of $\sigma$ for $\ell=2$. Right panel: Relative amplitude of the first four echoes compared to the amplitude of the transient. The points represent the simulated double/single wall pairs used in our analysis, as in Fig.~\ref{fig:echo_sigma}. At small widths, the transient decreases faster than the amplitudes of the echoes as $\sigma$ becomes smaller.} 
\end{figure*}

\begin{figure}[t!]
\centering
\includegraphics[width=.45\textwidth]{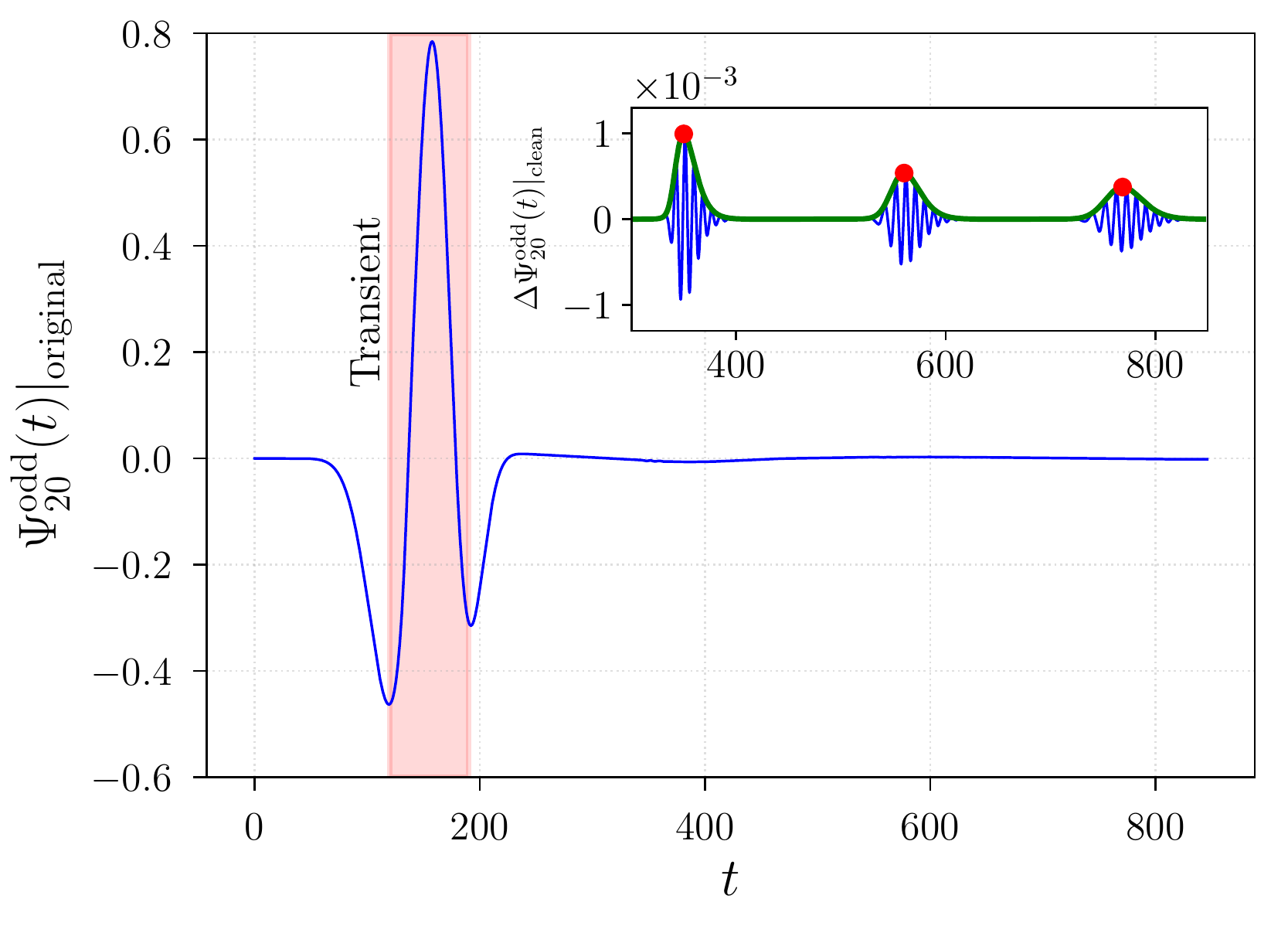}
\caption{\label{fig:echo_long} Asymptotic solution of an ingoing Gaussian pulse with $\sigma=16~r_g$ after its dispersion by a Morris-Thorne wormhole with a longer throat (with a length of 100~$r_g$). The amplitude of the transient is three orders of magnitude larger than the first echo, showing that the echoes do not have a generically large magnitude.}
\end{figure}

Frequency dependent reflection and transmission coefficients can be calculated by studying a scattering problem with a single potential wall, as we noticed in subsection \ref{subsec:bh}, this is simply achieved by doing an algebraic inversion of the tortoise coordinate definition in \eqref{eq:tortoise2}: here the inverted function is evaluated in $x-r_g$ instead of $|x|-r_g$. The definitions of the reflectivity and transmissivity coefficients for an individual potential wall remain the same as in \eqref{eq:ref_and_trans}

\begin{equation}
R(\omega)\equiv \frac{||\tilde{\Psi}_{\mathrm{ref}}^{\mathrm{odd}}(\omega)||}{||\tilde{\Psi}_{\mathrm{inc}}^{\mathrm{odd}}(\omega)||}~,~T(\omega)\equiv  \frac{||\tilde{\Psi}_{\mathrm{trans}}^{\mathrm{odd}}(\omega)||}{||\tilde{\Psi}_{\mathrm{inc}}^{\mathrm{odd}}(\omega)||},
\label{eq:odd_ref_and_trans}
\end{equation}
where we compute the one dimensional Fourier transforms of the incident $\tilde{\Psi}_{\mathrm{inc}}^{\mathrm{odd}}(\omega)=\mathcal{F}[\Psi^{\mathrm{odd}}_{\mathrm{bh}}(x,0)]$, reflected $\tilde{\Psi}_{\mathrm{ref}}^{\mathrm{odd}}(\omega)=\mathcal{F}[\Psi^{\mathrm{odd}}_{\mathrm{bh}}(+\infty,t)]$ and transmitted $\tilde{\Psi}_{\mathrm{trans}}^{\mathrm{odd}}(\omega)=\mathcal{F}[\Psi^{\mathrm{odd}}_{\mathrm{bh}}(-\infty,t)]$, where the label (bh) stands for the solutions of the scattering problem of \eqref{eq:wave_tensor_wh} with a single potential barrier. These single barrier solutions are not only necessary for the study of the potential cavity, but also to clean up the low frequency (high $\sigma$) solutions, since in those scenarios it is not simple to determine the amplitude of the echoes. All of the aforementioned definitions are also applicable for $\Psi^{\mathrm{even}}$. 

In the left panel of Fig.~\ref{fig:RT_reconst}, we show the reflection and transmission coefficients as functions of the frequency, noticing that the two curves intersect at $R^2=T^2=0.5$, as expected. The identity $R^2+T^2$ is approximately satisfied. As a next step of our analysis, we reconstruct the Fourier transform of the asymptotic pulse shown in Fig.~\ref{fig:odd_sigma_small}.  We employed the definition of the geometrical optics approximation in \eqref{eq:reconst}, applied up to $i=0$, in the Fourier transform of the Gaussian incident wavepacket in order to obtain the reconstructed profile in the right panel of Fig.~\ref{fig:RT_reconst}. The signal reconstructed using the geometrical optics approximation provides a better representation of the total reflected pulse as the ingoing wavelet gets wider, and therefore, it has more power in lower frequencies. 

Motivated by the drastic change in the amplitudes of the echoes seen in Figs.~\ref{fig:odd_sigma_small} and \ref{fig:odd_sigma_large}. We now explore the dependence of the amplitude of each individual echo with the width of the incident Gaussian pulse. To do so, we follow the same procedure explained by the end of subsection \ref{subsec:wh}: we construct a logarithmic grid in $\sigma$, centered at $\sigma_{\mathrm{DW}}=\sqrt{27}r_g/2\ell$ and spaced in intervals of $\sqrt{2}\sigma_{\mathrm{DW}}$. In addition to this, we define the variable $\Delta\Psi^{\mathrm{odd}}_{\mathrm{clean}}\equiv\Psi^{\mathrm{odd}}_{\mathrm{original}}-\Psi^{\mathrm{odd}}_{\mathrm{bh}}$ in order to clean the solutions from backscattering effects coming from the potential tails, which complicate the task of determining the amplitudes of the echoes with high $\sigma$/low frequency. Once the solutions are clean, the most effective way to find the maxima of each echo is by calculating the local maxima of the Hilbert envelope for the clean signal. In this case, the Hilbert envelopes are the green curves in the upper corner of Figs.~\ref{fig:odd_sigma_small} and \ref{fig:odd_sigma_large} and the maxima are the red dots on top of each curve. 
This labor is even more computationally expensive than in the scalar case, not only because we are solving the scattering problem for two systems -- one with a single potential barrier and another with the potential cavity -- but also we are now working with the two polarizations (i.e., the even and odd solutions). Our results of the amplitude analysis in Fig.~\ref{fig:echo_sigma_odd} show the existence of a value of $\sigma$ maximizing the amplitude of the echoes. This is compatible with the notion of a band of widths/frequencies in which the echoes have sufficient amplitude to be measured. 

It is reasonable to consider the length of the throat for a Morris-Thorne wormhole as a parameter modulating the echoes solutions. Thus, we rerun our simulation to consider the scattering of a Gaussian pulse with $\sigma=16~r_g$ by a wormhole with a larger throat (with a length of $100~r_g$, which is more than two times larger than the cavity used in the previous configuration). In the main panel of Fig.~\ref{fig:echo_long}, we show that the transient has a slightly larger magnitude: this is consistent with an increase of the wall reflectivity at lower frequencies seen in the left panel of Fig.~\ref{fig:RT_reconst}. Each of these potential barriers has the same shape of the potential barrier as in the scenario depicted in Fig.~\ref{fig:evenodd_potentials}, it is, therefore, still possible to find a frequency ``sweet spot'' for each wall. In the upper corner of Fig.~\ref{fig:echo_long}, we find that the amplitude of the echoes is three orders of magnitude smaller than the transient, after removing all backscattering effects from the signal. In the same figure, we notice that the time separation between echoes coincides with the time elapsed after two internal reflections, being greater than the time breach between echoes seen in Figs.~\ref{fig:asympt_two_sigmas}, \ref{fig:odd_sigma_small} and \ref{fig:odd_sigma_large}.

\subsection{Echoes and gravitational wave polarimetry}
\label{sec:Subsec_Polar}

\begin{figure}[t!]
\centering
\includegraphics[width=.45\textwidth]{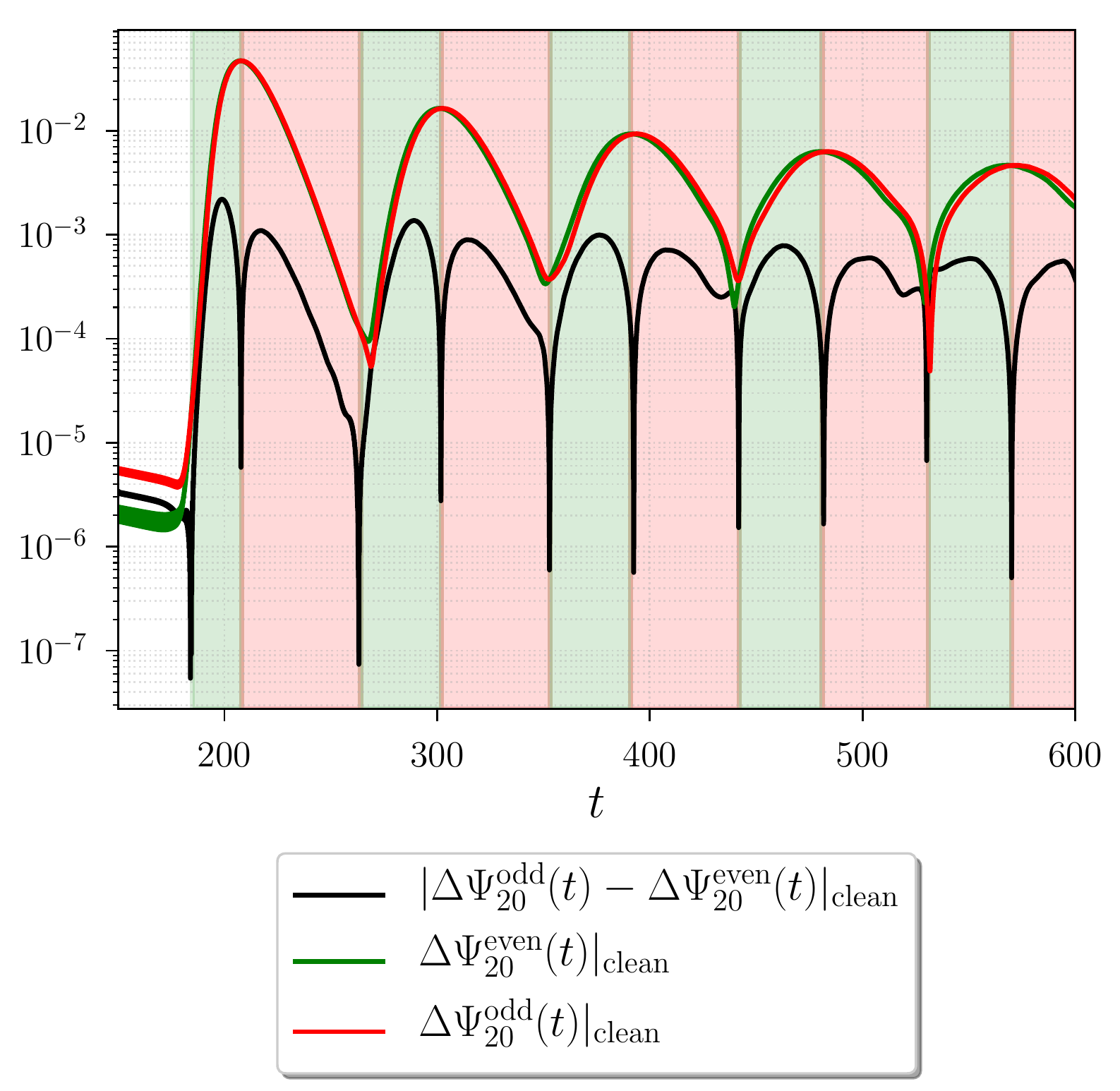}
\caption{\label{fig:pol} Since the phases of the asymptotic solutions are random, we compare the Hilbert envelopes of the two clean solutions for echoes. The outgoing envelopes have a small relative phase shift due to the difference in the potentials, inducing a net even $(h_+)$ polarization in the intervals shaded in green, and an odd $(h_{\times})$ polarization in the intervals shaded in red. Therefore, the cavity is analog to an optically active medium changing the polarization of the ingoing Gaussian wavepacket.}
\end{figure}

The small difference in the Zerilli (even) and Regge-Wheeler (odd) potentials is shown in the right panel of Fig.~\ref{fig:evenodd_potentials} has a particular effect in the outgoing waves. In order to illustrate it, we will just work with the even and odd quadrupole signals in the equatorial plane. In the case of a generic spherical mode with equal contributions from $\Psi^{\mathrm{odd}}_{20}$ and $\Psi^{\mathrm{even}}_{20} $, which are the first nontrivial contributions to \eqref{eq:hplus} and \eqref{eq:hx}, we notice that the two polarizations are reduced to

\begin{eqnarray}
h_+(x,t)=\frac{C_{\theta}}{r(|x|)}\Psi^{\mathrm{even}}_{20}(x,t),\nonumber\\
h_{\times}(x,t)=\frac{C_{\theta}}{r(|x|)}\Psi^{\mathrm{odd}}_{20}(x,t),\label{eq:pols}
\end{eqnarray}
where $C_{\theta}$ is a constant coming from the normalized spherical harmonics evaluated at $\theta=\pi/2$. Considering $\sigma=0.6495 r_g$, the parameters of the cavity used in sections \ref{sec:scalar} and \ref{sec:tensor} and the same initial Gaussian pulses for $\Psi^{\mathrm{even}}_{20}(x,t)$ and $\Psi^{\mathrm{odd}}_{20}(x,t)$.  As can be noticed in Fig.~\ref{fig:pol}, we show that the difference between the potentials induces a relative phase between the odd and even solutions, generating an outgoing wave with a net polarization oscillating from even $(h_+)$ to odd $(h_{\times})$ and only visible after the transient. It is interesting to notice that these small effects are present even when the target is spherically symmetric. Spin-orbit coupling between the spin-2 gravitational pertrubations and the angular momentum in Kerr-like solutions might enhance the polarization effects.

\section{Conclusions}\label{sec:conclusions}
In this paper, we studied the scattering of a test scalar and tensor wavepacket on a Morris-Thorne wormhole. Using a Gaussian pulse as an incident initial condition, we showed the time dependent scattering solution of the quadrupole scalar and tensor spherical modes both inside and outside the potential cavities in Figs.~\ref{fig:sol_n_flux_wh} and \ref{fig:even_evolve}, which reflect and transmit throughout the cavity. Furthermore, after finding the transmission and reflection coefficients of the cavities in Fig.~\ref{fig:RT_scalar} and in the left panel of Fig.~\ref{fig:RT_reconst}, we used the geometrical optics approximation to reconstruct the shape of the Fourier transformed asymptotic solutions in Fig.~\ref{fig:rec_scalar} and in the right panel of Fig.~\ref{fig:RT_reconst}. We find that the reconstructed shape of the spectrum is accurate, without showing, however, the QNM peaks.

In this paper, we show that in general, the echoes do not have a large amplitude as we can see directly in the left panel of Figs.~\ref{fig:asympt_two_sigmas}, \ref{fig:odd_sigma_large} and \ref{fig:echo_long}, where we also observe that the amplitude of the echoes does not decay exponentially in time. In addition to this, we found that there is a band of preferred frequencies (and the widths of the corresponding ingoing Gaussian signals) where the amplitude of the echoes is maximized. Such a frequency band is centered around the ``sweet spot'' in which the coefficients of transmissivity and reflectivity overlap, and it is precisely where the QNMs peaks are squeezed in. We extended our analysis to find the range in which the width of the incident pulses maximize the amplitude of the first four echoes, and how large is their amplitude compared to the transient. For small widths, this ratio could be as large as 0.15, we should notice, however, that the amplitude of the transient gets also suppressed in this range. In Figs.~\ref{fig:echo_sigma} and \ref{fig:echo_sigma_odd}, we found that for low widths of the ingoing signal, as the widths become smaller, the transient decreases faster than the amplitude of the echoes. In the study of the gravitational wave scattering by a Morris-Thorne wormhole, we find small differences between the Regge-Wheeler and the Zerilli effective potentials, as depicted in the right panel of Fig.~\ref{fig:evenodd_potentials}. As it is visible in Fig.~\ref{fig:pol}, such a difference modifies the polarization of any ingoing wave in a way analog to the dispersion across an optically active medium, opening the possibility of studying gravitational wave polarimetry.   

\begin{acknowledgments}
We would like to thank Alex Zucca, Levon Pogosian and Michael Desrochers for their time and their valuable comments and discussions. This project was partly funded by the Discovery Grants program of the Natural Sciences and Engineering Research Council of Canada. JG is supported by the Billy Jones Graduate Scholarship, granted by the Physics Department at SFU. This research was enabled in part by support provided by WestGrid (\url{www.westgrid.ca}) and Compute Canada Calcul Canada (\url{www.computecanada.ca}).
\end{acknowledgments}
    
\bibliography{bibliography.bib}

\end{document}